\documentclass[11pt,a4paper]{article}

\usepackage[T1]{fontenc}
\usepackage{lmodern}
\usepackage[a4paper,textwidth=468pt,vmargin=25mm]{geometry}
\usepackage{amsmath,amssymb}
\usepackage{graphicx}
\usepackage{booktabs,makecell,array,longtable}
\usepackage[font=small]{caption}
\usepackage[numbers,sort&compress]{natbib}
\usepackage[hidelinks]{hyperref}

\newcommand{\Aeven}{\Delta_{\mathrm{even}}}
\newcommand{\Sodd}{\Delta_{\mathrm{odd}}}

\title{How market volatility responds to a preceding price trend: a
lag-resolved measurement in cryptocurrencies and equities}

\author{Gülşah Bıçkı and Taylan Akdoğan\thanks{Corresponding author.
Email:
\href{mailto:taylan.akdogan@ozyegin.edu.tr}{taylan.akdogan@ozyegin.edu.tr}.
ORCID:
\href{https://orcid.org/0000-0002-0646-4637}{0000-0002-0646-4637}.}
\\[0.5ex] \normalsize Özyeğin University, İstanbul, Türkiye}

\date{}

\begin{document}

\maketitle

\begin{abstract}
We quantify the response of market volatility to a preceding price trend
as a function of lag at minute resolution. The asymmetry of the
response, i.e., the excess volatility following a fall relative to a
rise of the same magnitude, is investigated. The sample consists of
trades in three of the largest cryptocurrencies on a single exchange and
in three large US stocks and the S\&P~500 exchange-traded fund between
2020 and 2026. For the cryptocurrencies the asymmetry is positive,
matching the sign of the leverage effect. It follows the trend: it is
greatest at the first measurable lag, 20 minutes, and weak at negative
lags, where the trend follows the volatility. At those lags a scan over
window lengths resolves a small component of opposite sign that has not
yet been disentangled from what the statistic itself generates. For
Bitcoin and Ether the asymmetry decays to one half by a lag of roughly
an hour and a half. Decomposition under time reversal divides the
asymmetry into a $T$-even part and a $T$-odd part. Only the latter
carries an arrow of time. For Bitcoin and Ether the $T$-odd part is
detected. At the first lag it is between 0.4 and 0.9 of the $T$-even
part at windows of 10 and 20 minutes. At 20 minutes this is the ratio
produced by a purely one-sided response in this statistic: at that
window the response is consistent with being wholly one-sided in time.
None of the equities exhibits a detectable $T$-odd part at 20 minutes in
the straight-line fit over 2.5 trend scales. This fit is the shared
instrument of comparison. The index fund's asymmetry depends on that
fit: over the full range it is equally large at negative and positive
lags but of opposite sign, forming a mirror in lag whose two sides have
distinct origins; over the inner trends the positive side is four times
larger and the negative side is a quarter of it. The fund reacts to the
sign of a move more than to its magnitude. Its return–variance
correlation displays an arrow of time; the cryptocurrencies’ asymmetry
is independent of the shapes of the fits, and Bitcoin’s and Ether’s of
their ranges. On simulated and shuffled data without coupling, the
analysis yields zero with errors of the correct size. It also recovers
the time scale of a one-sided response embedded in a simulated market.

\medskip\noindent
\textbf{Keywords:} leverage effect, volatility asymmetry, response
function, time reversal, high-frequency data, cryptocurrency
\end{abstract}

\section{Introduction}
\label{sec:introduction}

A fall in a stock price is followed by greater volatility than a rise of
the same magnitude, and this leverage effect is a stylized fact of
financial returns \cite{cont2001}. Black \cite{black1976} observed it in
daily stock returns and suggested the explanation after which it is
named: a price fall increases the firm’s debt in relation to equity and
hence the risk of the share; he himself remarked that the effect seemed
too large to be explained by leverage alone. Christie
\cite{christie1982} tested this explanation and identified leverage as
the primary, though likely not the sole, cause. Schwert
\cite{schwert1989} reported that it accounts for only a small portion of
the temporal variation in market volatility. A second explanation,
volatility feedback, acts in the reverse direction: when expected
volatility increases, the price must drop for the stock to deliver the
higher return required by risk, so an increase in volatility causes a
fall \cite{campbell1992,bekaert2000}. The asymmetry is embedded in
standard volatility models, including the exponential GARCH of Nelson
\cite{nelson1991} and the GJR-GARCH of Glosten, Jagannathan and Runkle
\cite{glosten1993}. Its conventional representation is the news impact
curve of Engle and Ng \cite{engle1993}, the conditional variance plotted
against the signed shock. The effect reaches beyond the models, skewing
the distributions of returns over longer horizons and thereby the prices
of options \cite{bouchaud2001}.

The two explanations differ in temporal direction. In the former the
price move precedes the volatility response; in the latter, expected
volatility precedes the price move. A measurement that resolves lag in
both directions can distinguish between them. Such a measurement has
been carried out at daily lags. Bouchaud, Matacz and Potters
\cite{bouchaud2001} computed the correlation between a return and
squared returns at a subsequent time for hundreds of stocks and for
indices. The correlation decays over roughly ten days for indices and
roughly fifty days for individual stocks, with greater amplitude for the
indices. It vanishes at negative lags, where the squared return precedes
the return: the effect is one-sided in time. Their retarded volatility
model \cite{bouchaudpotters2003} and stochastic volatility models with
noise correlated with the price \cite{perello2003,perello2004} replicate
this one-sided exponential decay. For indices Reigneron, Allez and
Bouchaud \cite{reigneron2011} attributed the greater amplitude to the
correlation between stocks. This correlation increases following a fall.
When the daily volatility of US stocks is represented as a quadratic
function of past returns, the same pattern emerges: the component that
depends on the sign of past returns is weak, decays over weeks and is
near zero at negative lags, whereas the symmetric component accounts for
most of the feedback \cite{chicheportiche2014,blanc2014}. More
generally, financial time series are not invariant under time reversal:
volatility at long horizons drives volatility at shorter horizons more
than vice versa \cite{lynch2003,zumbach2009}. The one-sided leverage
correlation is another manifestation of the symmetry breaking. At
intraday resolution the measurement is more challenging. Bollerslev,
Litvinova and Tauchen \cite{bollerslev2006} employed five-minute
S\&P~500 futures returns and observed the correlation between a return
and the absolute returns following it to be negative from lag zero,
strongest at lag zero and decaying over at least five days. In the
reverse direction their correlations scatter about zero.
A{\"\i}t-Sahalia, Fan and Li \cite{aitsahalia2013} established that
naive high-frequency estimates of the leverage effect are biased toward
zero by discretization, by error in spot volatility and by
microstructure noise. Patton and Sheppard \cite{patton2015} demonstrated
with realized semivariances that future volatility is linked much more
strongly to the variation of past negative intraday returns than to that
of positive ones. Consistent estimators of the effect at lag zero have
since been devised. Wang and Mykland \cite{wang2014} estimate the
covariation of price with its variance, with and without microstructure
noise. Kalnina and Xiu \cite{kalnina2017} estimate the corresponding
correlation; A{\"\i}t-Sahalia et al.\ \cite{aitsahalia2017} isolate the
component arising from jumps and report it to be the larger one in the
Dow Jones index.

Cryptocurrencies offer a natural setting in which to study asymmetry in
the absence of a firm. There is no underlying firm and no balance sheet,
so the leverage explanation cannot apply. Any observed asymmetry would
call for an alternative cause. Their markets operate continuously, so a
lag axis at minute resolution is uninterrupted by closed-market gaps.
Prices are determined mainly by speculative demand, with a substantial
proportion of retail traders and leveraged derivatives. With daily data
the results are mixed. A comparison of GARCH models for Bitcoin favored
a two-component model lacking an asymmetry term \cite{katsiampa2017}. An
inverse asymmetry, with positive shocks increasing volatility more than
negative ones, was documented for daily Bitcoin returns prior to the
crash of December 2013, and none afterwards \cite{bouri2017}, and for
most of twenty currencies with a threshold GARCH \cite{baur2018} and for
three of four, Bitcoin among them, with a smooth-transition GARCH
\cite{cheikh2020}. A stochastic volatility model with long memory fitted
to 224 currencies instead yields the stock-like sign in each of them
\cite{phillip2018}. A correlation of this inverse sign, an anti-leverage
effect, had been observed previously at daily lags in the Chinese
indices of the 1990s, where the return--volatility correlation was
estimated lag by lag \cite{qiu2006,shen2009}. For Bitcoin and Ether a
threshold GARCH detects no asymmetry at all \cite{baur2018}, though a
quantile measure in the same study does reveal the stock-like pattern
for Bitcoin. GARCH, GJR and rational GARCH models fitted to daily
Bitcoin returns detect none either \cite{takaishi2018}. Daily data
cannot resolve intraday dynamics. If volatility responds to a price move
within hours, a daily estimator averages over the response and captures
little. The mixed daily results may reflect this limitation. The
one-minute statistics of Bitcoin, including its tails, correlations and
multifractality, have been analyzed and shown to approach those of the
foreign-exchange market as the market matured
\cite{takaishi2018,drozdz2018,watorek2021}. The data source is
important: parameters of a volatility model fitted to Bitcoin differ
beyond their uncertainty between an exchange and a coin-ranking site,
and fewer than half of cryptocurrency papers since 2017 were found to
employ correct data \cite{alexander2020}; prices on different exchanges
diverged from one another by more than the spread, by one to two orders
of magnitude in 2017 and 2018 \cite{makarov2020}. At intraday resolution
the return--volatility correlation of Bitcoin has been estimated lag by
lag with negative lags included: at two-minute sampling on one exchange
it is negative at positive lags, with a slow decay fitted by a power
law, and weak at negative lags \cite{takaishi2020}; at five-hour
sampling the leverage correlation has the same one-sided shape
\cite{desousa2021}. Models fitted at higher frequencies produce mixed
signs: inverted and diminishing over the years in a threshold GARCH of
six- and twelve-hour returns \cite{takaishi2021}, inverted in a panel of
five-minute data with a daily leverage term \cite{brini2024}, stock-like
after 2020 in asymmetric GARCH fits of five-minute returns of seven
currencies, though weak for Bitcoin \cite{ali2025}, and without a
systematic sign in a one-minute stochastic volatility model
\cite{nakakita2025}. The sign and one-sidedness of Bitcoin’s intraday
correlation are thus established \cite{takaishi2020}, as are those of
the index at five-minute resolution \cite{bollerslev2006}. This paper
corroborates both. What it contributes is a single instrument applied to
multiple cryptocurrencies and equities alike, with the trend as the
regressor, negative lags as a quantitative control, the parts even and
odd under time reversal separated, and a scale-free amplitude tracked
over the years.

This paper quantifies how market volatility responds to a preceding
price trend lag by lag at minute resolution for three cryptocurrencies
and, applying the same instrument, for three large US stocks and the
S\&P~500 exchange-traded fund. The result is regarded as a market
response function in the sense in which the price response to a trade is
defined \cite{bouchaud2004}. The impulse is the trend of the log price
over a 20-minute window, the response is the log variance over a
subsequent 20-minute window, and the lag between the two windows is the
variable of interest. The quantity under study is the asymmetry of the
response, i.e., the excess volatility following a fall relative to a
rise of the same magnitude; a positive asymmetry corresponds to the
direction of the leverage effect. The asymmetry is derived from fits of
log variance against trend on each side of zero. These fits constitute a
news impact curve resolved in lag. Realized volatility conditional on
the preceding trend was quantified by Zumbach in foreign exchange
\cite{zumbach2010}: trending prices increase it and mean-reverting
prices reduce it, regardless of trend direction. The asymmetry measured
here is the sign-dependent component of such conditioning. The symmetric
component, the increase in volatility following a move of either sign,
is not interpreted, as it mainly reflects the co-occurrence of large
moves and volatile periods.

Three design features underlie the measurement. First, lags are measured
in both the negative and positive directions. For negative lags the
trend window is positioned after the volatility window. A response to
the trend is one-sided in time, whereas a coupling without temporal
direction is not. The negative lags enable the two to be distinguished.
Second, neighboring lags are almost redundant. The unit for each error
is therefore a ten-week block of data, long relative to all time scales
in the response, rather than a single lag. Third, the entire analysis
was tested on simulated and shuffled data without coupling before
application to markets. Procedures that failed those tests were
abandoned.

For Bitcoin, Ether and Solana the asymmetry is positive and follows the
trend; for Bitcoin and Ether it decays to one half by a lag of roughly
an hour and a half, and at 20 minutes the part that carries an arrow of
time bears, within errors, the same ratio to the part that does not as
in a purely one-sided response embedded in a simulated market. In
equities no such one-sided part is detected at 20 minutes in the
straight-line fit over the full range of trends; in that fit the index
fund’s asymmetry is mirrored in lag, with the two sides having distinct
origins. The mirror depends on the range of trends covered by the fit.

\section{Data}
\label{sec:data}

The trades are drawn from the consolidated feed of Massive, previously
Polygon.io. For each trade the feed provides its price, size, time stamp
and exchange, and for equities its sale conditions and correction
indicator. Table~\ref{tbl:data} shows the assets, periods, filters and
the data that pass them.

\paragraph{Cryptocurrencies}
We use Bitcoin (BTC), Ether (ETH) and Solana (SOL). We take only trades
from a single exchange, Coinbase, the largest in the feed over the study
period. This is a methodological choice: prices across exchanges differ
by slight offsets at the trade level, and combining them introduces
symmetric noise into every window and smears the lag structure; a single
exchange yields a single price series. Cryptocurrencies trade around the
clock. The lag axis is therefore unbroken. The data cover 2020-01-20 to
2026-05-18 for Bitcoin and Ether. For Solana they extend from
2021-06-21, the first Monday following its listing on the exchange, to
2026-06-15. Prior to 2020 the exchange’s tape in the feed is sparser and
is not used.

\paragraph{Equities}
We use Apple (AAPL), Tesla (TSLA), Microsoft (MSFT) and the SPDR
S\&P~500 exchange-traded fund (SPY; the index fund hereafter). Their
trades are drawn from the consolidated tape of US exchanges and used in
regular trading hours only: 9:30 to 16:00 New York time on full trading
days. Holidays and half-days ahead of some holidays are omitted. A pair
of windows must lie within a single session of six and a half hours.
Equities are therefore measured out to a lag of roughly 260 minutes, 280
for the index fund, whereas cryptocurrencies extend to 600.

\paragraph{Epochs}
Data are divided into ten-week epochs starting on Mondays. Each epoch is
analyzed separately: 33 epochs for Bitcoin, Ether and the equities, 26
for Solana (Table~\ref{tbl:data}). The epoch is the unit for all errors
(Section~\ref{sec:errors}); four-week epochs reproduce the
cryptocurrencies’ amplitudes to within approximately 15\%
(Section~\ref{sec:robustness}). The first epoch, 2020-01-20 to
2020-03-30, includes the March 2020 crash, Bitcoin’s collapse on March
12 and 13 and the equity trough on March 23. It is retained; omission of
this epoch serves as a robustness check (Section~\ref{sec:robustness}),
not an exclusion.

\paragraph{Filters}
All trades from the exchange in the period are kept for the
cryptocurrencies, while equities undergo three filters. First, at fetch
every trade with a non-zero correction indicator is discarded. The
feed’s published condition table specifies for each sale condition
whether it updates the consolidated last price. Trades with a condition
that does not update are dropped: average-price, contingent, cash and
next-day trades and prior-reference and out-of-sequence reports, among
others. Also removed are reports that are not executions at the time
stamped (e.g., cancels, corrections, errors). So are sold-out and
stock-option trades, two conditions without a documented update rule
that we observed to be off the market. Two conditions that do not update
the last price are retained: odd lots, which make up roughly half of all
trades and are nearly always at the market, and extended-hours trades,
which fall outside the session and are removed by the third filter. On
three sample weeks in March 2020 and March 2024 the first filter
eliminated 0.1 to 1\% of regular-hours trades and 79 to 90\% of prints
more than 0.5\% off the market. Over the entire period and over all
hours of the tape it discarded 0.8\% (SPY) to 2.0\% (MSFT) of trades
(Table~\ref{tbl:data}); corrections account for 0.002 to 0.004\% of
trades. The remaining discarded trades carry one of the dropped
conditions. The share increases over time from 0.1 to 0.7\% of an
epoch’s trades in 2020 to 1.6 to 4.6\% in 2026. Second, one-trade spikes
are discarded: a trade that jumps by more than 1\% from the last kept
trade and whose successor reverts to within 20\% of the jump. This
eliminated 8,750 trades of the index fund and 284,188 of Tesla
(Table~\ref{tbl:data}), mainly single-share odd-lot prints a few percent
off the market. Third, trades outside the session are discarded.
Cleaning is consequential: on the raw tape single off-market prints
doubled the variance of 9.2\% of the index fund’s 20-minute windows in
the first quarter of 2020; after cleaning, the residual prints would do
so for 0.1 to 0.4\%. The cryptocurrency tape is not cleaned: its
infrequent off-market trades revert within two seconds of a price mark
(Section~\ref{sec:response-function}).

\paragraph{The clock and the grid}
The trades of an epoch are condensed into a grid of one-minute cells in
UTC, each storing additive moments. The trades are then discarded. Every
window is a sum of its cells; Section~\ref{sec:response-function}
describes what a cell contains. A return spanning a gap of more than
five minutes is discarded; this eliminated fewer than 0.01\% of the
returns in each asset. For the cryptocurrencies 99.2 to 99.7\% of the
minutes of the period contain a trade. The New York clock defines the
equities' session and the daily cycle of Section~\ref{sec:robustness};
for the cryptocurrencies, which have no session, the day is the UTC day.

\begin{table}[htp]
\centering
\caption{The data. The fourth column shows the fraction of equities’
trades, over all hours and the entire period, dropped by the fetch
filter (corrections and nonstandard sale conditions). Trades, trades
outside the session and trades kept are given in millions after that
filter; spikes are the one-trade spikes eliminated, expressed in trades.
Cryptocurrencies trade at all hours; equities are restricted to regular
hours, 9:30 to 16:00 New York time. The 33 epochs cover 2020-01-20 to
2026-05-18. Solana’s 26 epochs cover 2021-06-21 to 2026-06-15. The final
column is the count of instants sampled at the 20-minute lag over all
epochs; every epoch of every asset is fitted and pooled there.}
\label{tbl:data}
\scriptsize
\begin{tabular}{@{}llrrrrrrrr@{}}
\toprule
asset & market & epochs & \makecell[r]{dropped at\\the fetch, \%} & trades & \makecell[r]{outside\\the session} & spikes & kept & kept, \% & \makecell[r]{samples\\at 20 min} \\
\midrule
BTC  & Coinbase, all hours      & 33 & --   &   933.8 & --    & --      &   933.8 & 100.0 & 165,704 \\
ETH  & Coinbase, all hours      & 33 & --   &   747.8 & --    & --      &   747.8 & 100.0 & 165,005 \\
SOL  & Coinbase, all hours      & 26 & --   &   338.7 & --    & --      &   338.7 & 100.0 & 130,034 \\
AAPL & US tape, regular hours   & 33 & 1.78 & 1,017.6 &  50.3 &  36,995 &   967.3 &  95.1 &  29,962 \\
TSLA & US tape, regular hours   & 33 & 0.85 & 1,696.2 & 141.3 & 284,188 & 1,554.6 &  91.7 &  29,962 \\
MSFT & US tape, regular hours   & 33 & 2.04 &   629.4 &  29.0 &  22,020 &   600.4 &  95.4 &  29,962 \\
SPY  & US tape, regular hours   & 33 & 0.78 & 1,018.4 &  45.7 &   8,750 &   972.6 &  95.5 &  29,962 \\
\bottomrule
\end{tabular}
\end{table}

\section{Method}
\label{sec:method}

\subsection{The response function}
\label{sec:response-function}

\paragraph{The grid}
Trades in an epoch are condensed into the one-minute cells introduced in
Section~\ref{sec:data}. For variance the log price is marked at the
first trade arriving at least two seconds after the preceding mark. Each
cell stores the sum of squared log returns between successive marks and
the time covered by those returns. For trend, marks are at least ten
seconds apart. Each cell stores the sums necessary for a straight-line
fit of log price against time. Both sets of sums are additive over
cells. The variance and trend of any window of whole minutes therefore
follow exactly from sums over its cells, and the original trades are no
longer required once the grid is assembled.

\begin{figure}[tp]
\centering
\includegraphics[width=\textwidth]{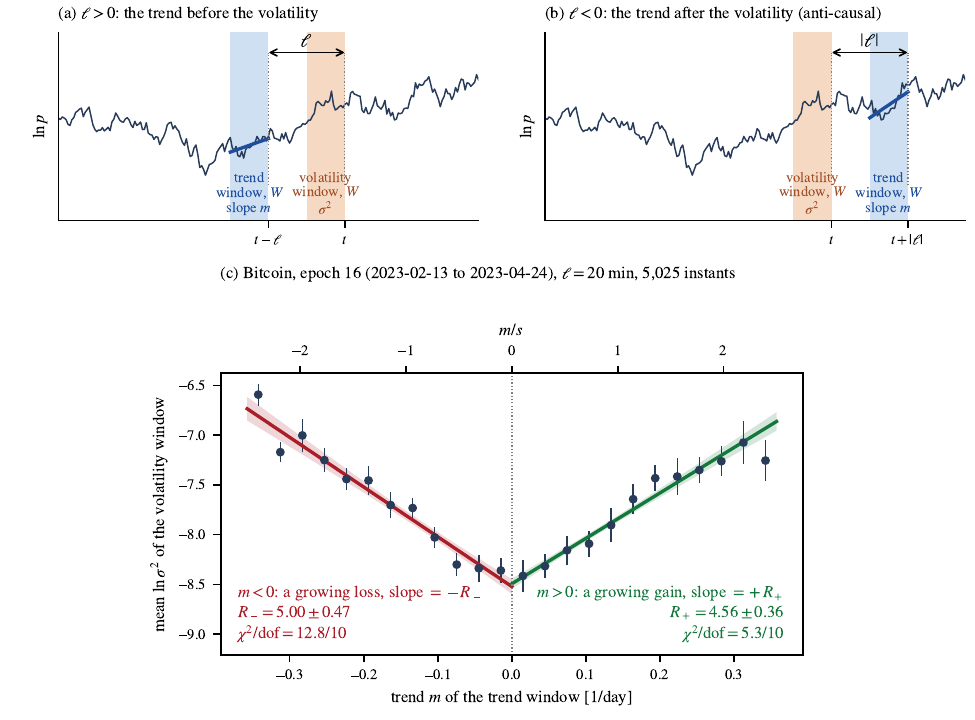}
\caption{The measurement. Top: a log-price trajectory with a trend
window of length $W = 20$~min and a volatility window of the same
length, displayed for a positive lag $\ell$ (trend preceding volatility;
at $\ell = W$ the windows are contiguous) and for a negative lag (trend
following volatility: the anti-causal side). Lag is defined as the
distance between the ends of the two windows. Bottom: one epoch of
Bitcoin (February to April 2023) at $\ell = 20$~min, mean log variance
of the volatility window plotted against the trend of the trend window
in trend bins, with a straight line fitted on each wing; the upper axis
shows the trend in units of its scale $s$. Each line stretches from the
outer edge of the outermost bin to the origin. Bars indicate the
dispersion of each bin’s mean across day-block replicas introduced in
Section~\ref{sec:errors}, against which $\chi^2$ is evaluated; the band
is the fit’s one-sigma band scaled to the replicas’ slope errors
(Appendix~\ref{app:error-model}). The slopes yield the responses to a
growing loss and a growing gain, $R_- = -c_1^-$ and $R_+ = +c_1^+$; the
asymmetry $\Delta = \tfrac12 (R_- - R_+)$ is positive when the down wing
is steeper. From the asymmetry at $\pm\ell$ the $T$-even part
$\Aeven(\ell) = \tfrac12[\Delta(\ell) - \Delta(-\ell)]$ and the $T$-odd
part $\Sodd(\ell) = \tfrac12[\Delta(\ell) + \Delta(-\ell)]$ are derived
(Section~\ref{sec:time-reversal}).}
\label{fig:measurement}
\end{figure}

\paragraph{The trend, the volatility and the lag}
Let $W = 20$~min. This value was set prior to analysis; the scan over
other windows in Section~\ref{sec:robustness} was carried out
subsequently as a check. For an instant $t$ the volatility window is
$[t-W, t)$. The trend window at lag $\ell$ is $[t-\ell-W, t-\ell)$
(Fig.~\ref{fig:measurement}). The trend $m$ is the slope of the straight
line fitted to the log price over the trend window, in units of 1/day.
It is the window’s return with parabolic weights on its steps. It
therefore incorporates every price in the window; for a random walk the
two correlate at 0.91, and in the correlation checks of
Section~\ref{sec:robustness} the return yields the same sign in Bitcoin
and Ether, after the move rather than before it. The volatility $v$ is
the logarithm of the realized variance of the volatility window: the sum
of squared log returns between its marks, divided by the time they
cover, in units of 1/day, so that a random walk produces the same value
for any window length. Instants advance by $W$. Volatility windows
therefore tile the record without overlap. Lag ranges from $-120$ to
$+600$ minutes in steps of 20, and it is measured between the ends of
the two windows, which is also the distance between their starts. For
$\ell \ge W$ the trend window precedes the volatility window: the causal
side. For $\ell \le -W$ it follows it: the anti-causal side, where a
response to the trend is impossible, although the statistic need not
vanish there (Section~\ref{sec:time-reversal}). At $\ell = W$ the
volatility window starts where the trend window ends. This lag, the
first lag, is the shortest at which the response can be measured. It is
also the lag reported. At $\ell = 0$ the two windows coincide. This lag
is skipped: the variance of the window feeds into its own trend, so the
measurement there does not represent a response. Sampling takes place
only when both windows are contained within the epoch and, for equities,
within the trading session. The causal lags for equities therefore stop
at approximately four and a half hours (Appendix~\ref{app:error-model}).
A cryptocurrency epoch yields roughly 5,000 instants per lag, whereas an
equity epoch yields roughly 900 (Table~\ref{tbl:data}).

\paragraph{The wing fits}
In every epoch and at every lag the samples $(m, v)$ are binned by
trend. The trend scale $s$ for the epoch is 1.4826 times the median
absolute deviation of $m$, i.e., the robust standard deviation. The
interval from $-2.5\,s$ to $+2.5\,s$ is partitioned into 24 equally wide
bins for cryptocurrencies and 12 for equities, which have roughly one
fifth of the samples; samples outside the interval are excluded, as are
bins with fewer than 30 samples (15 for equities). Within each bin the
mean of $v$ is calculated with its standard error; the average is taken
of the logarithm, not the logarithm of the average. A straight line,
\begin{equation}
  v = c_0 + c_1\, m ,
  \label{eq:wing}
\end{equation}
is then fitted by weighted least squares to the bin means versus the bin
centers separately on each side of zero; the two sides are termed the
down wing ($m < 0$) and the up wing ($m > 0$). The fits are separate: no
bin straddles zero and neither wing is folded onto the other. Expressed
against the magnitude of the trend, the two slopes are
\begin{equation}
  R_- = -c_1^- , \qquad R_+ = +c_1^+ ,
  \label{eq:wing-responses}
\end{equation}
the change in log variance per unit size of a prior loss and of a prior
gain, and from these we form
\begin{equation}
  \Delta(\ell) = \tfrac12\,[R_-(\ell) - R_+(\ell)] , \qquad
  \bar R(\ell)  = \tfrac12\,[R_-(\ell) + R_+(\ell)] .
  \label{eq:odd-even}
\end{equation}
$\Delta$ is the asymmetry of the response. It is positive when
volatility grows more after a fall than after a rise of the same
magnitude. This is the sign of the leverage effect. $\bar R$ is the
common response to a trend of either sign. In the terminology of the
news impact curve \cite{engle1993} the two wings match its two sides,
with the trend over $W$ as the shock and the log variance of the next
window as the impact, resolved lag by lag; $\Delta$ is one half the
difference of their slopes.

\paragraph{Why the common response is not interpreted}
A large $\bar R$ does not entail coupling. Large trends of either sign
appear in volatile periods, and volatility persists. Conditioning on a
large trend thus selects volatile states even without any directional
price effect. Data free of such coupling produce a large positive
$\bar R$ with $\Delta = 0$ (Section~\ref{sec:nulls}). We therefore
report and interpret $\Delta$ only, employing $\bar R$ as a control.

\subsection{Time reversal}
\label{sec:time-reversal}

Viewed backwards in time a rise becomes a fall and a trend window that
precedes the volatility window follows it, so time reversal maps
$\Delta(\ell)$ to $-\Delta(-\ell)$. For $\ell > 0$ we decompose the
asymmetry into
\begin{equation}
  \Aeven(\ell) = \tfrac12\,[\Delta(\ell) - \Delta(-\ell)] , \qquad
  \Sodd(\ell) = \tfrac12\,[\Delta(\ell) + \Delta(-\ell)] .
  \label{eq:time-reversal}
\end{equation}
$\Aeven$ is the $T$-even part, with $T$ standing for time reversal: it
is invariant under reversal of time. $\Sodd$ is the $T$-odd part: it
flips sign and alone carries an arrow of time. A response that follows
the trend and is absent before it obeys $\Delta(-\ell)=0$, so
$\Sodd=\Aeven$. A coupling without temporal direction obeys
$\Delta(-\ell)=-\Delta(\ell)$, the mirror image, so $\Sodd=0$ regardless
of the magnitude of $\Aeven$. A component that is equal on both sides,
$\Delta(-\ell)=\Delta(\ell)$, as would be generated by a contemporaneous
link spread to both sides by volatility persistence, is $T$-odd alone: a
$T$-odd part by itself does not signal a response. Interpretation
depends on the profile across lags. The decomposition contributes no
information beyond the two sides: with $r = \Delta(-\ell)/\Delta(\ell)$,
$\Sodd/\Aeven = (1+r)/(1-r)$. These are statements about the coupling;
the statistic itself produces a ratio under one even for a purely
one-sided response. This is because such a response skews the trend
distribution and the wings read that skewness at negative lags by an
amount quantified by the positive control in Section~\ref{sec:nulls} at
the 20-minute window; this reading stems from large trends and almost
disappears over the inner 1.5 trend scales. The two parts are computed
per epoch from the scale-free asymmetry defined in
Section~\ref{sec:quantities}, $\Delta_{s,e}(\ell)$ and
$\Delta_{s,e}(-\ell)$, at the six lags from 20 to 120 minutes available
on both sides, and then averaged over epochs. They are therefore
scale-free too and are comparable across epochs and assets; the interval
for the ratio $\Sodd/\Aeven$ is derived by resampling the epochs
(Section~\ref{sec:errors}).

\subsection{Errors and pooling}
\label{sec:errors}

\paragraph{Within an epoch}
Samples within a single epoch are not independent: adjacent instants
share a volatility state and volatility displays long memory. Each
sample is labeled by day. The error of every per-epoch quantity is
estimated by a block bootstrap over days. The 70 days of the epoch are
resampled in blocks of five consecutive days, repeated 200 times. Each
replica reruns the binning and the two wing fits; the dispersion of
$\Delta_e$ across replicas defines its error. Per-epoch values for which
the fit or bootstrap fails are excluded and exclusions are tallied per
lag (Appendix~\ref{app:error-model}).

\paragraph{Across epochs}
The ten-week epochs form the independent unit of the study. Asymmetry at
each lag and the scale-free amplitude defined in
Section~\ref{sec:quantities} are pooled across epochs as
inverse-variance weighted means. The pooled errors are enlarged by
$\sqrt{\chi^2/\mathrm{dof}}$ whenever epoch scatter surpasses that
permitted by the per-epoch errors, following the scale-factor rule of
the Particle Data Group \cite{pdg2024}. Alongside each pooled value an
epoch-block bootstrap is computed. It resamples epochs in blocks of two
consecutive epochs 400 times and ignores per-epoch errors entirely; the
two error estimates agree (Section~\ref{sec:crypto-response}).
Quantities computed per epoch from multiple lags — the $T$-even and
$T$-odd parts, band averages and the pooled decay curve — are reported
as unweighted means over epochs with the standard error of the mean. In
Bitcoin and Ether the epoch mean and the inverse-variance pooled
estimate agree within errors (Section~\ref{sec:crypto-response}); in the
index fund they do not, and both are presented
(Section~\ref{sec:equities}). Intervals for any quantity fitted to a
curve are derived from a bootstrap of the full analysis over epochs:
epochs are resampled, the curve is re-pooled and the fit is rerun, 1,500
times for the decay (Section~\ref{sec:quantities}).

\paragraph{Joint tests}
A joint test for deviation of a curve from zero is carried out across
lags with $Q = \sum_\ell [\Delta(\ell)/\sigma(\ell)]^2$ over the 30
causal lags and with the same statistic over the six anti-causal lags
from $-120$ to $-20$ minutes. The second test does not examine
one-sidedness: even for a purely one-sided response this statistic
rejects zero (Table~\ref{tbl:nulls}). $Q$ is not evaluated against a
$\chi^2$ distribution. Neighboring lags are almost redundant: adjacent
windows share most of their data, and the volatility of an epoch moves
all its lags together. Inferred from per-epoch residuals
(Appendix~\ref{app:error-model}), the effective number of independent
lags is roughly three of the 30 causal lags for cryptocurrencies, three
to five of the equities’ 13 or 14, and fewer than three of the six
anti-causal lags. The p-value of $Q$ is therefore calibrated by
resampling epochs under the null hypothesis; the lowest value that can
be reported is 0.0025, an upper bound. The same redundancy governs
errors for any quantity fitted to the curve: an error that presumes
independent lags is too small by factors of up to several hundred for a
fitted relaxation time (Appendix~\ref{app:error-model}). A model-free
check complements the pooled asymmetry at 20 minutes: the number of
epochs with positive $\Delta_e$, with its binomial p-value. Since epochs
differ in precision the count serves as a check rather than an estimate.

\subsection{The quantities quoted}
\label{sec:quantities}

\paragraph{The lag}
The asymmetry is reported at the first lag, 20 minutes. The response
decays over the subsequent hours. An average over a band of lags is
therefore band-dependent and dilutes the first lag; it does not measure
the response. Band averages are shown only in robustness checks
(Section~\ref{sec:robustness}).

\paragraph{The scale-free amplitude}
$\Delta$ is a slope per unit trend. The typical magnitude of the trend
increases with volatility. $\Delta$ therefore decreases as volatility
rises, and epochs with different volatility levels are not comparable
using $\Delta$. The scale-free amplitude
\begin{equation}
  \Delta_s = 2\, s\, \Delta ,
  \label{eq:scale-free}
\end{equation}
with $s$ standing for the trend scale of the epoch, is the difference in
log variance between a fall of one trend scale and a rise of one trend
scale, as long as the two wings intersect at zero trend. It is the
logarithm of a variance ratio and is dimensionless: $\Delta_s = 0.07$
signifies that variance following a typical fall is 7\% greater than
following a typical rise. The slopes do not capture a discontinuity
between the two wings at zero trend. We do not expect such a step in
cryptocurrencies: a market cannot differentiate a vanishingly small fall
from a vanishingly small rise. Any discontinuity between the fitted
lines would therefore need a response concentrated near zero trend,
which would modify the amplitude when the fit is restricted to the inner
1.5 trend scales or extended with a quadratic term
(Section~\ref{sec:robustness}). Bitcoin’s and Ether’s amplitudes stay
unchanged, whereas the index fund’s changes, reflecting its response to
the sign of the move (Section~\ref{sec:mirror}). $\Delta_s$ is invariant
to rescaling of epoch volatility and therefore enables comparison across
epochs, years and assets. It is computed per epoch and lag and pooled in
the same way as $\Delta$; the time-reversal parts $\Aeven$ and $\Sodd$
defined in Section~\ref{sec:time-reversal} are obtained from it per
epoch and averaged over epochs. Their subscripts therefore indicate
symmetry rather than sign.

\paragraph{The two channels}
Repeating the same analysis with variance rather than its logarithm
defines the linear channel. There the asymmetry is normalized by $c_0$,
the mean baseline of the two wings, producing a relative change. The log
channel is primary because it weights all windows equally, whereas
variance is dominated by the most volatile windows. Both channels are
reported (Section~\ref{sec:robustness}).

\paragraph{The half-decay lag}
The reported time scale is the first lag at which the pooled curve drops
to one half of its 20-minute value, located by linear interpolation
between lags. The curve pertains to $\Delta$; within a given epoch
$\Delta_s$ equals $\Delta$ multiplied by a factor constant across lags.
Each epoch’s decay is therefore the same for both. It is defined only
when the 20-minute value lies at least three standard errors from zero,
because a noise curve can halve by chance. Its interval is the 16th to
84th percentile across the 1,500 epoch-bootstrap replicas. No functional
form is assumed. It is not a relaxation time: it is anchored to the
20-minute value and combines decay with any floor and with the offset
induced by trend skewness at every lag (Section~\ref{sec:nulls}).

\paragraph{The assumed form of the decay}
We represent the decay by a single relaxation time along with a floor,
\begin{equation}
  \Delta(\ell) = b + a\, e^{-\ell/\tau} , \qquad \ell \ge 20~\text{min} .
  \label{eq:assumed-form}
\end{equation}
This corresponds to the linear response of mean-reverting volatility to
a brief impulse in trend. It is one-sided in time, thus
$\Sodd = \Aeven$. This functional form arises in standard asymmetric
volatility models and in stochastic volatility models with correlated
noise \cite{perello2003,perello2004} and is the form observed for the
leverage correlation of stocks and indices at daily lags
\cite{bouchaud2001}. The floor $b$ absorbs modes slower than the
ten-hour window, since a mode slower than a day is effectively constant
within it. The functional form is an assumption, and over 20 to 600
minutes the data cannot discriminate it from a power law with a floor
(Appendix~\ref{app:shape}). The fit is applied to the pooled curve, and
$\tau$ and $b$ derive their intervals from the bootstrap of the full
analysis over epochs. We term $\tau$ the model relaxation time; it is a
single-method quantity (Section~\ref{sec:time-scale}).

\subsection{The null tests as a gate}
\label{sec:null-gate}

Prior to encountering market data every estimator in this study was
tested on data in which the coupling between price direction and
volatility is zero by construction. Each estimator was required to
return zero with correctly sized errors. Four families of such null data
were created. Each family was written once to the trade cache with a
record of its construction and then passed through exactly the
configuration of its parent asset; only the ticker differs and the study
driver confirms this.
\begin{itemize}
  \item \emph{A simulated market without coupling,} three seeds. The log
  variance is a sum of five Ornstein--Uhlenbeck components with
  relaxation times ranging from eight minutes to sixty days; the log
  price proceeds in Gaussian steps with that variance, sampled
  independently of volatility innovations; trades arrive as a Poisson
  stream whose rate tracks the variance. This retains a realistic
  volatility process and trade flow while containing no link from price
  direction to volatility. The same market with a known one-sided
  response embedded, in which log variance reacts to prior normalized
  price steps with leverage sign and a single relaxation time of 90
  minutes, serves as the positive control, with three seeds
  (Section~\ref{sec:nulls}).
  \item \emph{The trade-level sign shuffle of Bitcoin,} three seeds. The
  sign of each inter-trade return is replaced by a fair coin toss and
  prices are recomputed. Timing, trade sizes, return magnitudes and
  volatility clustering are retained; the direction of every trade is
  erased.
  \item \emph{The minute shuffle of the index fund,} three seeds. One
  coin per minute multiplies all returns within that minute. The
  intra-minute path is thus either retained or mirrored in its entirety:
  bid–ask bounce, reverting prints, session structure, overnight gaps
  and the intraday volatility profile are all maintained, and only the
  net direction of each minute is randomized. A trade-level shuffle is
  unsuitable for a consolidated equity tape, where off-market prints
  must revert: on the raw tape of the index fund it increased two-second
  variance 24-fold, whereas the minute shuffle leaves it unchanged.
  \item \emph{The negate mirror of Bitcoin.} The sign of every return is
  negated, mirroring the path: every trend switches sign while variances
  are unaltered. This is not a zero null but a known transformation. The
  asymmetry must appear as the exact negative of Bitcoin’s, with the
  common response unchanged.
\end{itemize}
For both shuffles the expected value of $\Delta$ is exactly zero for any
estimator: flipping all coins simultaneously maps a shuffled stream onto
its mirror image, and the two are equally likely. Since the shuffles
retain the volatility process of the real data they also test the
errors: the $t$-values under the null must be distributed as standard
normal noise. This rule is strict. An estimator is applied to market
data only after it returns zero with correctly sized errors on data
without coupling and the exact negative under mirroring. Two estimators
failed this gate and were abandoned: a comparison of decay shape
(Appendix~\ref{app:shape}) and the first design of daily-cycle removal
(Section~\ref{sec:robustness}). Nulls were created for Bitcoin and the
index fund only (Section~\ref{sec:limits}); their results are presented
in Section~\ref{sec:nulls}.

\section{Results}
\label{sec:results}

\subsection{The response of the cryptocurrencies}
\label{sec:crypto-response}

\begin{figure}[tp]
\centering
\includegraphics[width=\textwidth]{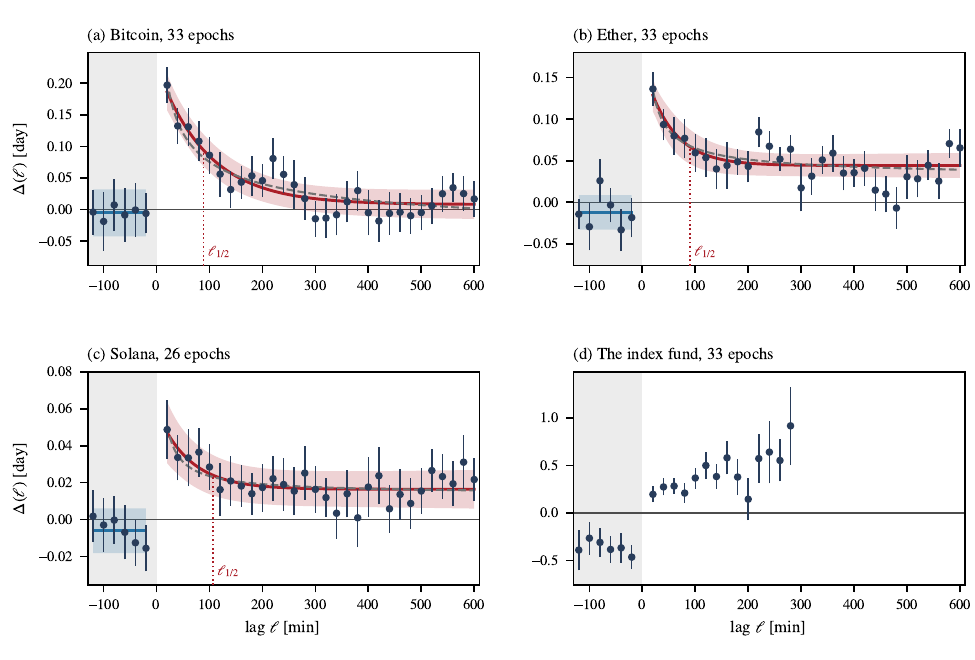}
\caption{The response as a function of lag. The asymmetry $\Delta(\ell)$
from the log channel, pooled across epochs with the epoch as the unit
(mean across epochs and its standard error at each lag), for (a)
Bitcoin, (b) Ether, (c) Solana and (d) the index fund; the three stocks
are displayed in Appendix~\ref{app:stocks}
(Fig.~\ref{fig:response-stocks}). The gray region indicates anti-causal
lags. For the cryptocurrencies the fits from
Section~\ref{sec:time-scale} are plotted: over anti-causal lags a
constant (blue) computed as each epoch’s mean over its anti-causal lags
averaged across epochs with its standard error; over causal lags the
exponential with floor $b + a\,e^{-\ell/\tau}$ (red), with its band
denoting the standard deviation of the fitted curve across the 1,500
epoch-bootstrap replicas, the power law with a floor (dashed) and the
half-decay lag $\ell_{1/2}$ (dotted). The index fund has no fits, since
its asymmetry is below zero at anti-causal lags
(Section~\ref{sec:equities}). Its points stop at 280~minutes: both
windows and the lag must fall within a single trading session of six and
a half hours, so instants become sparse as lag lengthens. Beyond
280~minutes too many epochs lose their fit for pooling to stay viable
(Section~\ref{sec:response-function}; 260~minutes for the stocks). Each
panel uses a separate vertical scale. The pooled values from
Section~\ref{sec:errors}, reported in the text, weight epochs by their
errors and deviate slightly from the points plotted here for Bitcoin and
Ether and more for the index fund (Section~\ref{sec:equities}).}
\label{fig:response}
\end{figure}

Figure~\ref{fig:response} plots the response of the three
cryptocurrencies as a function of lag. Table~\ref{tbl:main} lists the
values at 20 minutes. In Bitcoin and Ether the asymmetry is small at
negative lags, consistent with zero for this window (a shorter window
resolves it, Section~\ref{sec:robustness}), largest at the first causal
lag, 20 minutes, and decreases over the subsequent hours; in Ether it
tends toward a floor of roughly one third of its 20-minute value,
whereas in Bitcoin it tends toward roughly zero
(Section~\ref{sec:time-scale}). The constant fitted to anti-causal lags
is consistent with zero for all three cryptocurrencies:
$-0.005 \pm 0.037$ for Bitcoin, $-0.012 \pm 0.021$ for Ether and
$-0.006 \pm 0.012$ for Solana. At 20 minutes the asymmetry — the epoch
mean employed in Fig.~\ref{fig:response} and in the decay fits — is
$+0.197 \pm 0.028$ for Bitcoin and $+0.136 \pm 0.021$ for Ether. Pooling
epochs by inverse variance with the scale factor
(Section~\ref{sec:errors}), the joint test over the 30 causal lags
rejects zero at the lowest $p$-value the calibration can report,
$p < 0.0025$. The anti-causal test does not, with $p = 0.46$ and $0.58$.
The pooled estimate yields $+0.182 \pm 0.019$ and $+0.130 \pm 0.015$ at
20 minutes, consistent with the epoch mean: it up-weights the more
precise epochs, unlike the unweighted mean which treats all epochs
equally. The epoch-block bootstrap returns errors of 0.018 and 0.014
alongside scale-factor errors of 0.019 and 0.015. This confirms
consistency between the two error models of Section~\ref{sec:errors}.

Solana is weaker, with a 20-minute asymmetry of $+0.049 \pm 0.016$ and a
pooled estimate of $+0.029 \pm 0.013$; the causal test returns
$p = 0.035$ and the anti-causal test $p = 0.28$. Its response is
concentrated at short lags: beyond roughly 100 minutes the curve reaches
its floor (Fig.~\ref{fig:response}c). The sign across epochs is in line
with the pooled values: the 20-minute asymmetry is positive in 31 of 33
epochs for both Bitcoin and Ether, and in 19 of Solana’s 26 epochs
(Table~\ref{tbl:main}). Because epochs vary in precision, these counts
verify the sign of the pooled estimate without replacing its error.
Scale-free amplitudes from Table~\ref{tbl:main} are examined in
Section~\ref{sec:amplitude} and the equities’ rows in
Section~\ref{sec:equities}.

\begin{table}[htp]
\centering
\caption{Main results at 20 minutes for each asset. $\Delta$ is the
log-channel asymmetry given as the mean over epochs with standard error,
as shown in Fig.~\ref{fig:response}. The two $p$-values refer to joint
tests over causal lags (30 for cryptocurrencies, 13 or 14 for equities)
and over six anti-causal lags; both pool epochs by inverse variance with
the scale factor (Section~\ref{sec:errors}) and are calibrated by epoch
resampling. The lowest reportable value is 0.0025.
$\Delta_s = 2 s \Delta$ is the scale-free amplitude
(Eq.~\ref{eq:scale-free}), pooled across epochs by inverse variance with
scale-factor errors for both log and linear channels; the linear channel
drops epochs where the ratio is undefined because the baseline is close
to zero. There are 5 to 8 such epochs per asset and none for Solana. The
superscript on the log-channel values for Bitcoin, Ether and the index
fund indicates a one-sided systematic stemming from the spacing of
variance marks: the amplitude is greater by that amount per epoch when
realized variance is computed from one-minute returns rather than
two-second marks (Section~\ref{sec:robustness}); this was not evaluated
for Solana and the stocks. The final column reports the number of epochs
with positive $\Delta_e$ at 20 minutes and the one-sided binomial
$p$-value; because epochs vary in precision the count serves as a check
rather than an estimate.}
\label{tbl:main}
\scriptsize
\begin{tabular}{@{}lrrrrrrr@{}}
\toprule
asset & epochs & $\Delta$ [day] & \makecell[r]{causal\\$p$} & \makecell[r]{anti-causal\\$p$} & \makecell[r]{$\Delta_s$,\\log} & \makecell[r]{$\Delta_s$,\\linear} & \makecell[r]{epochs with\\$\Delta_e > 0$ ($p$)} \\
\midrule
BTC  & 33 & $+0.197 \pm 0.028$ & $< 0.0025$ & 0.46  & $+0.071 \pm 0.008^{\,+0.012}$ & $+0.130 \pm 0.026$ & 31 ($7 \times 10^{-8}$) \\
ETH  & 33 & $+0.136 \pm 0.021$ & $< 0.0025$ & 0.58  & $+0.066 \pm 0.008^{\,+0.014}$ & $+0.141 \pm 0.022$ & 31 ($7 \times 10^{-8}$) \\
SOL  & 26 & $+0.049 \pm 0.016$ & 0.035      & 0.28  & $+0.022 \pm 0.009$ & $+0.056 \pm 0.022$ & 19 (0.014) \\
\midrule
SPY  & 33 & $+0.196 \pm 0.083$ & $< 0.0025$ & 0.005 & $+0.026 \pm 0.015^{\,+0.025}$ & $+0.025 \pm 0.041$ & 20 (0.15) \\
AAPL & 33 & $+0.082 \pm 0.053$ & 0.005      & 0.10  & $+0.027 \pm 0.015$ & $+0.028 \pm 0.030$ & 21 (0.08) \\
TSLA & 33 & $+0.056 \pm 0.026$ & 0.080      & 0.30  & $+0.035 \pm 0.014$ & $+0.079 \pm 0.028$ & 22 (0.04) \\
MSFT & 33 & $+0.164 \pm 0.072$ & 0.017      & 0.24  & $+0.039 \pm 0.018$ & $+0.068 \pm 0.030$ & 22 (0.04) \\
\bottomrule
\end{tabular}
\end{table}

\subsection{Time reversal: the \texorpdfstring{$T$}{T}-odd and the \texorpdfstring{$T$}{T}-even part}
\label{sec:directed-part}

\begin{figure}[tp]
\centering
\includegraphics[width=\textwidth]{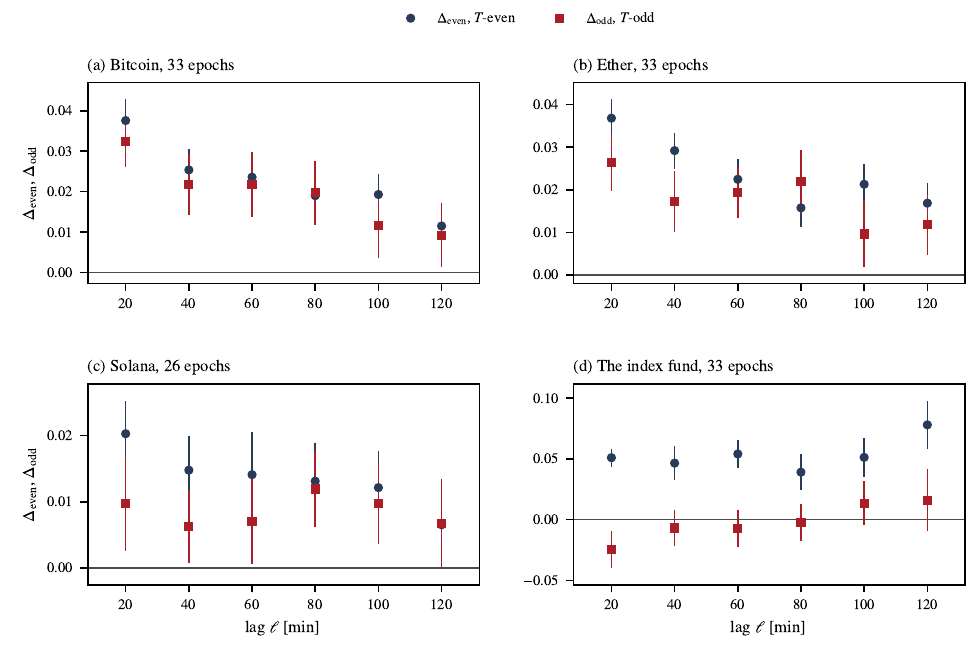}
\caption{Time reversal. The $T$-even part $\Aeven$ (circles) and the
$T$-odd part $\Sodd$ (squares) of the scale-free asymmetry $\Delta_s$,
Eq.~(\ref{eq:time-reversal}) evaluated per epoch, are plotted at the six
lags from 20 to 120 minutes, averaged across epochs with standard error
of the mean, for (a) Bitcoin, (b) Ether, (c) Solana and (d) the index
fund; the three stocks appear in Appendix~\ref{app:stocks}
(Fig.~\ref{fig:time-reversal-stocks}). A response absent before the
trend satisfies $\Sodd = \Aeven$. A mirror satisfies $\Sodd = 0$. Each
panel uses a separate vertical scale.}
\label{fig:time-reversal}
\end{figure}

Figure~\ref{fig:time-reversal} plots the $T$-even part $\Aeven$ and the
$T$-odd part $\Sodd$ from Eq.~(\ref{eq:time-reversal}) at six lags from
20 to 120 minutes, computed per epoch from the scale-free asymmetry
$\Delta_s$ (Section~\ref{sec:quantities}). Table~\ref{tbl:time-reversal}
tabulates the values. In Bitcoin and Ether the $T$-odd part is of
comparable magnitude to the $T$-even part. At 20 minutes,
$\Aeven = +0.038 \pm 0.005$ and $\Sodd = +0.032 \pm 0.006$ for Bitcoin,
and $\Aeven = +0.037 \pm 0.004$ and $\Sodd = +0.026 \pm 0.007$ for
Ether. The $T$-odd part is thus $5.1\sigma$ and $4.0\sigma$ away from
zero. Across the band $\Sodd/\Aeven$ is $+0.85$ for Bitcoin with a 68\%
interval 0.53–1.24, and $+0.76$ for Ether (0.49–1.05). Both intervals
contain one, the value expected for a response absent before the trend.
They exclude zero, the value for a mirror. Within these uncertainties,
at this window the response is one-sided in time. Its $T$-odd part is
$4\sigma$ to $5\sigma$ from zero and comparable in magnitude to the
$T$-even part. Solana is consistent with the same pattern but weaker:
$\Sodd = +0.010 \pm 0.007$ at 20 minutes and $\Sodd/\Aeven = +0.63$
(0.21–1.22). The ratio varies with window length
(Section~\ref{sec:robustness}), and at the first lag it equals 0.5, 0.9,
0.4 and 0.6 for Bitcoin and 0.4, 0.7, 0.3 and 0.0 for Ether at $W = 10$,
20, 40 and 60 minutes. The $T$-odd part is $2.6\sigma$ and $2.4\sigma$
from zero at 10 minutes and within $1.7\sigma$ at 40 and 60 minutes,
where its error roughly doubles; consequently the ratio is reliably
measured only at 10 and 20 minutes, where it ranges from 0.4 to 0.9. The
reported window yields the highest value. The $T$-odd part therefore
amounts to 0.4–0.9 of the $T$-even part where it is measured and is
comparable at 20 minutes. At 20 minutes this matches the ratio generated
by a purely one-sided response in this statistic: the positive control
in Section~\ref{sec:nulls}, a simulated market with a one-sided response
of Bitcoin’s amplitude and a 90-minute relaxation time and no backward
action, evaluated at the 20-minute window, yields $0.65 \pm 0.05$ at the
first lag and roughly 0.5 across the band. The control’s $T$-even and
$T$-odd parts lie within one to two standard errors of Bitcoin’s. The
20-minute ratios are therefore those expected for a one-sided response;
the control was not evaluated at the other windows, so the lower ratios
at 10 minutes lack calibration.

The index fund constitutes the opposite case. Its $T$-even part is
large, $\Aeven = +0.051 \pm 0.007$ at 20 minutes, whereas its $T$-odd
part is $\Sodd = -0.025 \pm 0.015$. Across the band
$\Sodd/\Aeven = -0.04$ with an interval $-0.29$ to $+0.26$, which
excludes one. Its asymmetry is mirrored in lag
(Section~\ref{sec:equities}). Removing the daily volatility cycle
(Section~\ref{sec:robustness}; Table~\ref{tbl:robustness}) leaves the
$T$-even parts unchanged. The $T$-odd parts of Bitcoin and Ether
decrease to roughly two thirds of the $T$-even part
($\Sodd/\Aeven = +0.62$ and $+0.61$ over the band; $+0.025 \pm 0.007$
and $+0.023 \pm 0.006$ at 20 minutes) but remain significant; the
difference is treated as a systematic uncertainty for the $T$-odd part.
Values for the stocks (Appendix~\ref{app:stocks},
Table~\ref{tbl:time-reversal-stocks}) are examined in
Section~\ref{sec:equities}.

\begin{table}[htp]
\centering
\caption{Time reversal. The $T$-even part $\Aeven$ and the $T$-odd part
$\Sodd$ of the scale-free asymmetry $\Delta_s$,
Eq.~(\ref{eq:time-reversal}) evaluated per epoch on $\Delta_{s,e}(\ell)$
and $\Delta_{s,e}(-\ell)$, are listed for the six lags common to both
sides, averaged across epochs with standard error of the mean, in units
of $10^{-2}$ (the unit of $\Delta_s$ in Table~\ref{tbl:main}); the final
column contains the ratio of the two band means over 20 to 120 minutes,
given as the median across epoch-bootstrap replicas with its 68\%
interval. A response absent before the trend satisfies $\Sodd = \Aeven$.
A mirror satisfies $\Sodd = 0$. The three stocks appear in
Appendix~\ref{app:stocks} (Table~\ref{tbl:time-reversal-stocks}).}
\label{tbl:time-reversal}
\scriptsize
\setlength{\tabcolsep}{4pt}
\begin{tabular}{@{}llrrrrrrl@{}}
\toprule
 & & \multicolumn{6}{c}{lag $\ell$ [min]} & \\
\cmidrule(lr){3-8}
asset & & 20 & 40 & 60 & 80 & 100 & 120 & $\Sodd/\Aeven$, 20--120 min \\
\midrule
BTC  & $\Aeven$ & $+3.8 \pm 0.5$ & $+2.5 \pm 0.5$ & $+2.4 \pm 0.6$ & $+1.9 \pm 0.6$ & $+1.9 \pm 0.5$ & $+1.1 \pm 0.5$ & $+0.85$ [+0.53, +1.24] \\
     & $\Sodd$ & $+3.2 \pm 0.6$ & $+2.2 \pm 0.8$ & $+2.2 \pm 0.8$ & $+2.0 \pm 0.8$ & $+1.2 \pm 0.8$ & $+0.9 \pm 0.8$ & \\
ETH  & $\Aeven$ & $+3.7 \pm 0.4$ & $+2.9 \pm 0.4$ & $+2.2 \pm 0.5$ & $+1.6 \pm 0.5$ & $+2.1 \pm 0.5$ & $+1.7 \pm 0.5$ & $+0.76$ [+0.49, +1.05] \\
     & $\Sodd$ & $+2.6 \pm 0.7$ & $+1.7 \pm 0.7$ & $+1.9 \pm 0.6$ & $+2.2 \pm 0.7$ & $+1.0 \pm 0.8$ & $+1.2 \pm 0.7$ & \\
SOL  & $\Aeven$ & $+2.0 \pm 0.5$ & $+1.5 \pm 0.5$ & $+1.4 \pm 0.6$ & $+1.3 \pm 0.6$ & $+1.2 \pm 0.6$ & $+0.7 \pm 0.6$ & $+0.63$ [+0.21, +1.22] \\
     & $\Sodd$ & $+1.0 \pm 0.7$ & $+0.6 \pm 0.6$ & $+0.7 \pm 0.6$ & $+1.2 \pm 0.6$ & $+1.0 \pm 0.6$ & $+0.7 \pm 0.7$ & \\
\midrule
SPY  & $\Aeven$ & $+5.1 \pm 0.7$ & $+4.6 \pm 1.4$ & $+5.4 \pm 1.2$ & $+3.9 \pm 1.5$ & $+5.1 \pm 1.6$ & $+7.8 \pm 2.0$ & $-0.04$ [$-0.29$, +0.26] \\
     & $\Sodd$ & $-2.5 \pm 1.5$ & $-0.7 \pm 1.5$ & $-0.7 \pm 1.5$ & $-0.2 \pm 1.5$ & $+1.3 \pm 1.8$ & $+1.6 \pm 2.6$ & \\
\bottomrule
\end{tabular}
\end{table}

\subsection{The amplitude over the years}
\label{sec:amplitude}

\begin{figure}[tp]
\centering
\includegraphics[width=\textwidth]{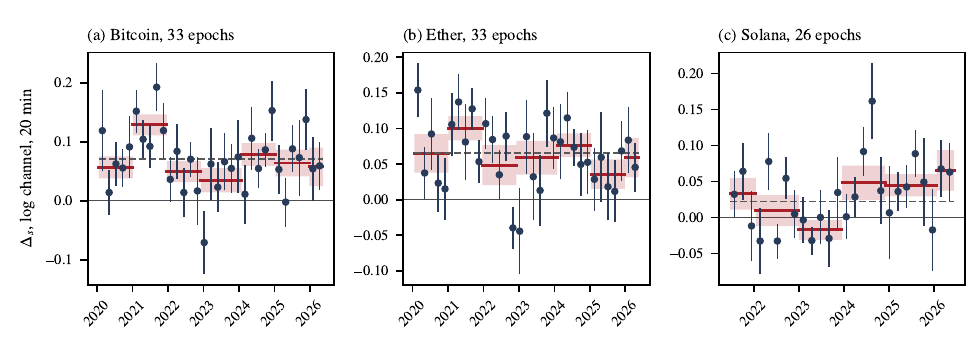}
\caption{Scale-free amplitude $\Delta_s$ of the log channel at 20
minutes, plotted epoch by epoch versus the epoch midpoint with error
bars for (a) Bitcoin, (b) Ether and (c) Solana. The segments represent
the pooled value for each calendar year, computed by inverse-variance
weighting with the scale factor (Section~\ref{sec:errors}). Each segment
is displayed with its error band over the epochs whose midpoint lies in
that year; the dashed line is the pooled value across all epochs
(Table~\ref{tbl:main}). Each panel uses a separate vertical scale.}
\label{fig:amplitude-by-year}
\end{figure}

Table~\ref{tbl:main} lists the scale-free amplitude from
Eq.~(\ref{eq:scale-free}) at 20 minutes. Bitcoin and Ether display the
same coupling, $\Delta_s = 0.071 \pm 0.008$ and $0.066 \pm 0.008$ in the
log channel: variance following a typical fall is roughly 7\% greater
than following a typical rise. Solana shows roughly one third of this
effect, $0.022 \pm 0.009$. Figure~\ref{fig:amplitude-by-year} plots the
amplitude epoch by epoch along with the pooled value per calendar year.
For Bitcoin and Ether the amplitude is positive in each year from 2020
to 2026. The two halves of the sample are consistent: $0.076 \pm 0.014$
versus $0.066 \pm 0.011$ for Bitcoin and $0.067 \pm 0.015$ versus
$0.063 \pm 0.010$ for Ether. A linear fit through the epochs is
consistent with zero slope, $-0.7\sigma$ and $-1.3\sigma$. Solana’s
effect emerges in the second half of its record, $0.049 \pm 0.011$
versus $0.002 \pm 0.011$ in the first, with a slope of $+2.2\sigma$.

The amplitude exhibits no systematic dependence on volatility level.
Regressing on the logarithm of the epoch’s baseline variance yields
slopes of $+0.022 \pm 0.011$ for Bitcoin, $+0.021 \pm 0.013$ for Ether
and $-0.034 \pm 0.013$ for Solana: two are positive and one negative,
with significances $2.0\sigma$, $1.6\sigma$ and $2.6\sigma$. The trend
scale itself scales with the square root of baseline variance, as
expected under pure rescaling. The slopes are $0.50 \pm 0.03$ for
Bitcoin and $0.53 \pm 0.02$ for Ether; Solana’s slope, $0.33 \pm 0.06$,
is lower. In the linear channel the amplitude is roughly double that in
the log channel for the three cryptocurrencies, $0.130 \pm 0.026$,
$0.141 \pm 0.022$ and $0.056 \pm 0.022$ (Table~\ref{tbl:main}); the
per-epoch difference is $2.3$ to $3.8\sigma$ from zero
(Section~\ref{sec:robustness}). The equities’ amplitudes are examined in
Section~\ref{sec:equities}.

\subsection{The time scale}
\label{sec:time-scale}

The cryptocurrency response decays over the hours following the trend
(Fig.~\ref{fig:response}). The model-free time scale from
Section~\ref{sec:quantities} is defined as the first lag at which the
pooled curve declines to half its 20-minute value. It is 89 minutes for
Bitcoin with a 16th–84th percentile interval 70–106, 90 minutes for
Ether (52–123) and 107 minutes for Solana (53–135): roughly one and a
half hours for Bitcoin and Ether (Table~\ref{tbl:time-scale}). The scale
is defined for all three assets because the 20-minute values lie
$6.9\sigma$, $6.6\sigma$ and $3.1\sigma$ from zero and at most 0.3\% of
replicas fail to halve. Solana’s interval is broad because its 20-minute
value is the least significant. The time scale is therefore reported for
Bitcoin and Ether. Windows of 20 to 60 minutes produce half-decay lags
of 86–120 minutes for Bitcoin and 90–100 minutes for Ether, and a
10-minute window resolves a faster component with a half-decay lag of
35–39 minutes (Section~\ref{sec:robustness}).

Under the assumed form of Eq.~(\ref{eq:assumed-form}), a single
relaxation time with a floor, the fit to the pooled curve returns
$\tau = 95$ minutes for Bitcoin (64–136), 53 minutes for Ether (24–157)
and 62 minutes for Solana (41–187). The intervals come from the
bootstrap of the full analysis over epochs. The floor is consistent with
zero for Bitcoin, $0.008$ ($-0.019$ to $0.029$), and amounts to roughly
one third of the 20-minute value for Ether and Solana, $0.044$ ($0.024$
to $0.056$) and $0.016$ ($0.004$ to $0.027$). Figure~\ref{fig:response}
plots the fits with their bands. The fit employs weighted least squares
over the 30 causal lags. It weights by per-lag standard errors with
$\tau$ profiled on a grid; the residual sum is not a $\chi^2$ because
lags are correlated, and no inference is drawn from it. With roughly
three effectively independent lags (Appendix~\ref{app:error-model}),
three parameters is the maximum the window can support. Models with more
parameters, such as two relaxation times or a stretched exponential,
were therefore not tried. $\tau$ is a single-method estimate. The second
method specified in advance, a likelihood pooled over per-epoch decays,
yields 248 minutes for Bitcoin with an interval 188–350, does not
resolve Ether and Solana, and fails its own consistency checks; it is
not used (Appendix~\ref{app:shape}). The functional form itself remains
an assumption. Over 20 to 600 minutes the data do not discriminate it
from a power law with a floor: in the bootstrap the exponential fits
better than the power law in 85\% of replicas for Bitcoin, 42\% for
Ether and 87\% for Solana, below the 90\% threshold set in advance for a
preference. The same test favors the exponential in 95–100\% of replicas
on data without coupling, where noise should impart no preference, so
the test is inconclusive (Section~\ref{sec:nulls};
Appendix~\ref{app:shape}). We therefore report the half-decay lag as the
time scale and $\tau$ as the parameter of the assumed model. The
half-decay lag is shape-free but not independent of the statistic: for a
one-sided response with Bitcoin’s amplitude the positive control yields
a half-decay lag of 69–74 minutes versus 82 minutes implied by its
90-minute relaxation time, short by a tenth to a sixth because trend
skewness depresses the curve by a constant offset at all lags; the fit
with a floor absorbs this offset and recovers $\tau$ within a tenth
(Section~\ref{sec:nulls}). The reported half-decay lag may therefore be
biased low by a comparable amount. $\tau$ provides a better
relaxation-time estimate if the assumed decay form holds. Over the inner
1.5 trend scales the control’s offset almost disappears and its
half-decay lag is 79–85 minutes, but the data’s own half-decay lags
there are too uncertain to report.

\begin{table}[htp]
\centering
\caption{The time scale of the cryptocurrency response in units of
$\Delta$. The 20-minute value is the epoch mean as plotted in
Fig.~\ref{fig:response}; the half-decay lag is the first lag at which
the pooled curve declines to one half of that value, determined by
linear interpolation between lags; $\tau$ and the floor $b$ are
parameters of the assumed form, Eq.~(\ref{eq:assumed-form}), fitted to
the pooled curve over 20 to 600 minutes. All intervals are the 16th to
84th percentile across the 1,500 epoch-bootstrap replicas. The second
method for $\tau$ specified in advance, a likelihood pooled over
per-epoch decays, yields 248 minutes (188–350) for Bitcoin, does not
resolve Ether and Solana, and fails its own checks; it does not
corroborate these values (Appendix~\ref{app:shape}).}
\label{tbl:time-scale}
\scriptsize
\begin{tabular}{@{}lrrrr@{}}
\toprule
asset & value at 20 min & half-decay lag [min] & $\tau$ [min] & floor $b$ \\
\midrule
BTC & $+0.197 \pm 0.028$ &  89 [70, 106] & 95 [64, 136] & $+0.008$ [$-0.019$, $+0.029$] \\
ETH & $+0.136 \pm 0.021$ &  90 [52, 123] & 53 [24, 157] & $+0.044$ [$+0.024$, $+0.056$] \\
SOL & $+0.049 \pm 0.016$ & 107 [53, 135] & 62 [41, 187] & $+0.016$ [$+0.004$, $+0.027$] \\
\bottomrule
\end{tabular}
\end{table}

\subsection{The equities}
\label{sec:equities}

Equities were examined with the same windows and estimator and behave
differently. Table~\ref{tbl:main} reports a positive asymmetry at 20
minutes for the three stocks, $1.5\sigma$ to $2.3\sigma$ each, with
scale-free amplitudes 0.027 to 0.039, i.e., two fifths to one half of
Bitcoin’s; causal tests produce $p = 0.005$ for Apple, 0.08 for Tesla
and 0.017 for Microsoft. The stocks’ anti-causal tests are passed. Yet
the $T$-odd part is not detected in any of the four assets under the
straight-line fit over 2.5 trend scales employed for all numbers
reported thus far; fits over alternative ranges are considered later in
this section. At 20 minutes $\Sodd$ is $-0.009 \pm 0.017$ for Apple,
$+0.013 \pm 0.016$ for Tesla, $+0.021 \pm 0.019$ for Microsoft and
$-0.025 \pm 0.015$ for the index fund (Tables~\ref{tbl:time-reversal}
and~\ref{tbl:time-reversal-stocks}). All lie within $1.7\sigma$ of zero.
This is the case for the index fund at every lag up to two hours.
Apple’s and Tesla’s band ratios, $\Sodd/\Aeven = +0.01$ and $+0.17$,
have 68\% intervals that contain zero and exclude one. Their 95\%
intervals contain both. Microsoft’s $T$-even part over the band is
consistent with zero while its $T$-odd part climbs to $+0.059 \pm 0.018$
at 80 minutes. This leaves the ratio unresolved and too noisy to
interpret. Non-detection does not mean absence. One-sided 95\% upper
limits on the band mean of $\Sodd$ are 0.022 for the index fund, 0.023
for Apple and 0.029 for Tesla. The band means for Bitcoin and Ether are
$+0.019 \pm 0.007$ and $+0.018 \pm 0.006$: a $T$-odd part considerably
larger than that of the cryptocurrencies is ruled out, but one of
similar size is not. For the index fund the 95\% interval for
$\Sodd/\Aeven$, $-0.49$ to $+0.67$, is compatible with a $T$-odd part up
to roughly two thirds of its $T$-even part.

Under this fit the index fund’s asymmetry is mirrored in lag
(Fig.~\ref{fig:response}d). At positive lags the pooled curve grows from
$+0.20 \pm 0.08$ at 20 minutes to approximately $+0.5$ past two hours
(the inverse-variance pooled estimate from the joint tests climbs from
$+0.07 \pm 0.07$ to roughly $+0.3$). The joint test over causal lags
rejects zero at $p < 0.0025$. At negative lags the asymmetry is of equal
magnitude with opposite sign: the constant over anti-causal lags is
$-0.36 \pm 0.14$. The anti-causal test rejects zero at $p = 0.005$. Lag
by lag the two sides agree within errors across the band, with the
negative side $1.1$ to $1.4$ times the positive between 40 and 80
minutes. The first lag breaks this pattern: there the negative side is
roughly three times the positive, $-0.076 \pm 0.019$ versus
$+0.026 \pm 0.014$ in scale-free units. The time-reversal decomposition
corroborates the mirror: the $T$-even part is $+0.051 \pm 0.007$ at 20
minutes and lies between $+0.04$ and $+0.08$ out to two hours, the
$T$-odd part is consistent with zero throughout and $\Sodd/\Aeven$ over
the band is $-0.04$ with a 68\% interval $-0.29$ to $+0.26$. In this
statistic the index fund is not a one-sided response of the sort seen in
cryptocurrencies; the nature of its mirrored sides is addressed in
Section~\ref{sec:mirror}. The fund’s return–variance correlation does
display an arrow of time in the leverage direction as documented by
Bollerslev, Litvinova and Tauchen \cite{bollerslev2006}; the wing
asymmetry, which also depends on the shape of later returns, does not. A
10-minute window shows the same mirror; at 40 minutes the negative side
is unresolved while the positive side is substantially larger
(Section~\ref{sec:robustness}).

The mirror pertains to the straight-line fit over 2.5 trend scales. The
fund’s findings depend on that choice whereas Bitcoin’s and Ether’s do
not (eighth check, Section~\ref{sec:robustness};
Table~\ref{tbl:robustness}). Over the inner 1.5 scales the fund’s
first-lag amplitude is $0.110 \pm 0.026$ versus $0.026 \pm 0.015$. In
scale-free units the two sides at 20 minutes over that range are
$+0.124 \pm 0.028$ and $-0.033 \pm 0.023$, the negative being roughly
one quarter of the positive; with a quadratic term per wing the slope at
zero trend yields an amplitude of $0.30 \pm 0.07$. The negative side
then ceases to be negative. The $T$-odd part at 20 minutes is
$-0.025 \pm 0.015$ over 2.5 scales, $+0.046 \pm 0.023$ over 1.5 scales
and $+0.21 \pm 0.06$ with the quadratic term; over 4 scales the mirror
survives. The three stocks shift in the same direction by one to three
standard errors, while Bitcoin’s and Ether’s amplitudes, negative sides
and $T$-odd parts remain unaltered under any of these variations. The
fund responds more to the sign of the move than to its magnitude
(Section~\ref{sec:mirror}). A slope fitted over a broad trend range
captures little of a response localized near zero trend. We therefore
present the fund’s first-lag asymmetry and $T$-odd part as limits
particular to the fit. We retain the 2.5-scale fit as the shared
instrument of comparison. The positive control in
Section~\ref{sec:nulls} shows how far each range departs from a
one-sided response. Over 2.5 scales a one-sided response yields
$\Sodd/\Aeven = 0.65 \pm 0.05$ at the first lag. The fund’s
$-0.49 \pm 0.31$ lies $3.6\sigma$ away. Over 1.5 scales, where the
control is almost clean at $0.88 \pm 0.06$, the fund’s $0.58 \pm 0.30$
is one standard error below it and two above a mirror: near zero trend
the fund’s numbers do not distinguish a one-sided response from a
mirror.

Its causal test over positive lags only is as significant as Bitcoin’s.
The mirror is not a data artifact: it persists after cleaning the tape
(Section~\ref{sec:data}), it is eliminated by a null that randomizes the
direction of each minute (Section~\ref{sec:nulls}) and it stays after
subtracting the intraday volatility profile, with
$\Aeven = +0.050 \pm 0.007$ and $\Sodd = -0.012 \pm 0.012$ at 20 minutes
(Table~\ref{tbl:robustness}). For the index fund the epoch mean in
Table~\ref{tbl:main}, $+0.20 \pm 0.08$ at 20 minutes, and the
inverse-variance pooled estimate from the joint tests, $+0.07 \pm 0.07$,
diverge more than for cryptocurrencies: the fund’s epochs vary widely in
precision and three epochs contribute half the pooled weight.

Our conclusion concerning equities applies only to this horizon: the
$T$-odd part is measured from 20 minutes to two hours, where both sides
of a lag stay within the session, and the asymmetry is evaluated out to
roughly four and a half hours. The leverage effect of stock indices at
daily lags is well established \cite{bouchaud2001}.

\subsection{The nulls}
\label{sec:nulls}

\begin{figure}[tp]
\centering
\includegraphics[width=\textwidth]{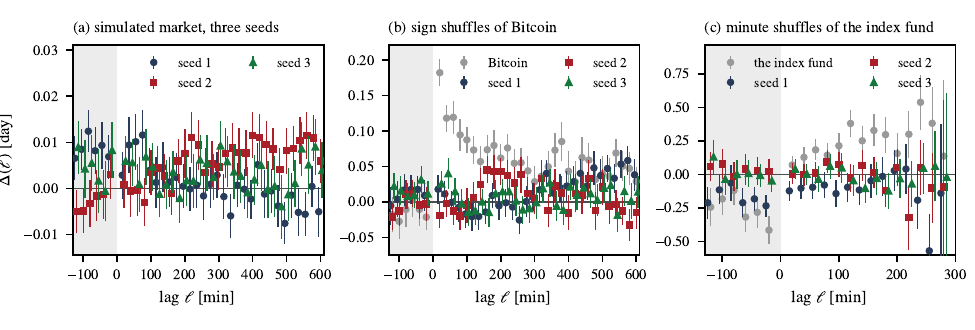}
\caption{The nulls. Asymmetry of the log channel versus lag, pooled over
epochs by inverse variance with the scale factor (the estimator used in
Table~\ref{tbl:main}) for three of the four families of data without
coupling (Section~\ref{sec:null-gate}): (a) the three seeds of the
simulated market; (b) the three sign shuffles of Bitcoin with Bitcoin’s
own curve plotted in gray; (c) the three minute shuffles of the index
fund with the index fund’s curve plotted in gray. The negate mirror of
Bitcoin, which reproduces Bitcoin’s curve with the sign exactly
reversed, is omitted (Table~\ref{tbl:nulls}). Seeds are shifted by a few
minutes so their error bars do not overlap. The gray region indicates
anti-causal lags, and each panel has its own vertical scale. In (c) the
lag axis terminates at the session boundary and the vertical scale is
determined by lags up to 200 minutes; beyond that few epochs survive the
censoring rule (Appendix~\ref{app:error-model}) and the error bars
extend off the scale.}
\label{fig:nulls}
\end{figure}

Section~\ref{sec:null-gate} sets out the four families of data without
coupling and the rule that an estimator is applied to markets only after
yielding zero with correctly sized errors and the exact negative under
mirroring. Figure~\ref{fig:nulls} plots the pooled asymmetry for three
of the families with the parent asset’s curve in the background.
Table~\ref{tbl:nulls} reports the values for all four. At 20 minutes the
nulls lie within $1.7\sigma$ of zero. Bitcoin is at $9.8\sigma$. No
causal joint test rejects zero: $p$-values span 0.09 to 0.94.
Anti-causal tests likewise pass with a single exception — the first seed
of the simulated market at $p = 0.035$, one of eighteen tests at the 5\%
level; its other two seeds pass at $p = 0.60$ and 0.27 with anti-causal
points scattered around zero. This indicates that the first seed’s tilt
is a fluctuation particular to that seed. Across all lags the highest
signal-to-error ratio on any null is 2.8, compared with 9.8 for Bitcoin
and 4.0 for the index fund. The scatter of epochs around each pooled
value, quantified as $\chi^2$ per degree of freedom per lag, has a
median of 0.9 to 1.5 over the nine zero nulls. Per-epoch errors are
therefore correctly sized. The sign shuffles retain Bitcoin’s volatility
process and produce zero where Bitcoin does not (Fig.~\ref{fig:nulls}b).
The negate mirror reproduces Bitcoin’s asymmetry with the sign reversed,
to $1.5 \times 10^{-12}$ in the pooled curve and $4 \times 10^{-11}$ in
every epoch and lag. It does not alter the common response.

The minute shuffle erases the index fund’s mirror. Its anti-causal test,
which rejects zero on the data at $p = 0.005$, is passed on all three
shuffles with $p = 0.05$, 0.93 and 0.87. The constant over anti-causal
lags drops from $-0.36 \pm 0.14$ to $-0.11 \pm 0.10$, $+0.03 \pm 0.11$
and $-0.01 \pm 0.09$. Only the net direction of each minute is
randomized; intra-minute paths, session structure, overnight gaps and
the intraday volatility profile are preserved. The mirror is therefore a
property of price direction within the day, not of tape structure. Two
estimators failed this gate and were abandoned. The comparison of decay
shapes favored the exponential on every null in 95 to 100\% of replicas,
where noise should produce no preference. The shape thus cannot be
inferred (Appendix~\ref{app:shape}). The first design of daily-cycle
removal, which also removed a trend profile, did not return zero on the
nulls; the design adopted in Section~\ref{sec:robustness} removes only
the volatility profile and passes all four families.

The positive control is a simulated market with an embedded one-sided
response: log variance reacts to past normalized price steps with
leverage sign and a single relaxation time of 90 minutes. There is no
backward action. Two gain levels were tried at the 20-minute window with
three seeds each: one matching Bitcoin’s first-lag amplitude,
$\Delta_s = 0.069$ to $0.079$, and one 2.4 times larger. The estimator
recovers the response and its time scale: the causal test rejects zero
in every seed, and the fit of Eq.~(\ref{eq:assumed-form}) yields
$\tau = 85$ to 98 minutes at Bitcoin’s amplitude and 87 to 91 minutes at
the larger amplitude, versus the 90 minutes specified. It does not yield
a zero negative side or $\Sodd = \Aeven$. At Bitcoin’s amplitude the
constant over anti-causal lags is $-0.017$ to $-0.027 \pm 0.005$, one
sixth to one quarter of the first-lag value with opposite sign, and the
anti-causal test rejects zero in every seed; $\Sodd/\Aeven$ at the first
lag is 0.56, 0.71 and 0.67 with errors of roughly 0.1, and 0.4 to 0.6
across the band. At the larger amplitude the negative side is one fifth
to one third and the ratio is roughly one half. The cause is the
statistic, not a coupling: a one-sided response magnifies steps after a
fall within the trend window, skewing the trend distribution. Because
volatility persists a positive trend of a given size implies higher
variance before it than a negative trend of the same size, which the
wing slopes interpret as asymmetry at negative lags. The half-decay lag
is biased low by the same mechanism, to 69–74 minutes versus 82 minutes
implied by the relaxation time, because the mechanism’s constant affects
every lag; the fit with a free floor absorbs it and so leaves $\tau$
intact. Table~\ref{tbl:nulls} reports the six seeds. For a purely
one-sided response at Bitcoin’s amplitude, $\Aeven = 0.044$ to $0.050$
and $\Sodd = 0.028$ to $0.034$. These values lie within one to two
standard errors of Bitcoin’s $+0.038 \pm 0.005$ and $+0.032 \pm 0.006$.
Over the inner 1.5 trend scales, the range employed in the eighth check
of Section~\ref{sec:robustness}, the three streams are almost clean:
$\Sodd/\Aeven$ at the first lag is 0.81 to 0.99, the negative side is
$-0.08$ to $+0.01$ of the first-lag value with the anti-causal test
passing, and the half-decay lag is 79 to 85 minutes. The statistic
registers trend skewness in large trends. Bitcoin’s amplitude does not
rise over the inner range as the control’s does, by roughly one third,
and its ratio over the full range, about 0.85, exceeds the control’s:
the mechanism is no larger in Bitcoin than in the control and may be
smaller.

\begin{table}[htp]
\centering
\caption{Summary of the nulls. Each family of data without coupling
(Section~\ref{sec:null-gate}) is presented under exactly its parent’s
configuration: the feature it removes, the asymmetry at 20 minutes
pooled by inverse variance with the scale factor, the estimator employed
for the joint tests, the calibrated joint tests over causal and
anti-causal lags, and the largest signal-to-error ratio at any lag.
Parents are shown for comparison. The negate mirror is not a zero null:
it must reproduce the parent’s values with sign reversed. It does so to
$1.5 \times 10^{-12}$ in the pooled curve and $4 \times 10^{-11}$ per
epoch and lag. The one-sided kernel is the positive control, not a zero
null: a response with a 90-minute relaxation time and no backward
action, at two amplitudes; it must display the response and it does; its
rejection of zero by the anti-causal test stems from the statistic’s
interpretation of the skewness induced in the trend by a one-sided
response (Section~\ref{sec:nulls}).}
\label{tbl:nulls}
\scriptsize
\begin{tabular}{@{}llrrrr@{}}
\toprule
data & removes & $\Delta$ at 20 min & causal $p$ & anti-causal $p$ & \makecell[r]{largest\\$|\Delta/\sigma|$} \\
\midrule
Bitcoin                & --                                        & $+0.182 \pm 0.019$ & $< 0.0025$ & 0.46  & 9.8 \\
simulated market, seed 1 & any link from the direction to the volatility & $+0.003 \pm 0.005$ & 0.56 & 0.035 & 2.7 \\
simulated market, seed 2 &                                           & $+0.001 \pm 0.005$ & 0.090 & 0.60 & 2.3 \\
simulated market, seed 3 &                                           & $+0.006 \pm 0.006$ & 0.62  & 0.27 & 1.7 \\
sign shuffle, seed 1   & the direction of every trade              & $+0.018 \pm 0.021$ & 0.087      & 0.69  & 2.8 \\
sign shuffle, seed 2   &                                           & $-0.019 \pm 0.018$ & 0.46       & 0.64  & 2.4 \\
sign shuffle, seed 3   &                                           & $+0.025 \pm 0.018$ & 0.63       & 0.33  & 1.8 \\
negate mirror          & nothing; every return flipped             & $-0.182 \pm 0.019$ & $< 0.0025$ & 0.46  & 9.8 \\
\midrule
one-sided kernel, seed 1 & a one-sided response, 2.4 times Bitcoin's & $+0.221 \pm 0.008$ & $< 0.0025$ & $< 0.0025$ & 28.5 \\
one-sided kernel, seed 2 &                                           & $+0.227 \pm 0.007$ & $< 0.0025$ & $< 0.0025$ & 32.1 \\
one-sided kernel, seed 3 &                                           & $+0.252 \pm 0.008$ & $< 0.0025$ & $< 0.0025$ & 30.0 \\
one-sided kernel, seed 1 & the same at Bitcoin's amplitude           & $+0.109 \pm 0.005$ & $< 0.0025$ & $< 0.0025$ & 20.4 \\
one-sided kernel, seed 2 &                                           & $+0.104 \pm 0.005$ & $< 0.0025$ & 0.005 & 19.3 \\
one-sided kernel, seed 3 &                                           & $+0.104 \pm 0.008$ & $< 0.0025$ & 0.005 & 12.3 \\
\midrule
the index fund         & --                                        & $+0.070 \pm 0.069$ & $< 0.0025$ & 0.005 & 4.0 \\
minute shuffle, seed 1 & the net direction of each minute          & $-0.123 \pm 0.074$ & 0.22       & 0.052 & 2.8 \\
minute shuffle, seed 2 &                                           & $+0.004 \pm 0.078$ & 0.77       & 0.93  & 1.4 \\
minute shuffle, seed 3 &                                           & $+0.040 \pm 0.077$ & 0.94       & 0.87  & 1.0 \\
\bottomrule
\end{tabular}
\end{table}

\subsection{Robustness}
\label{sec:robustness}

\begin{figure}[tp]
\centering
\includegraphics[width=\textwidth]{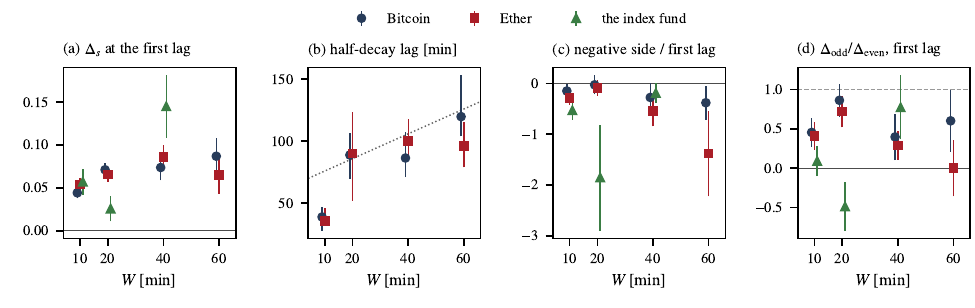}
\caption{Window-length scan: the plain runs at $W = 10$, 20, 40 and 60
minutes (sixth check; values in Table~\ref{tbl:robustness}). (a)
Scale-free amplitude $\Delta_s$ at the first lag, $\ell = W$. (b)
Half-decay lag for cryptocurrencies with 16th to 84th percentile
interval; the dotted line represents $W + \tau \ln 2$ for $\tau = 95$
minutes, where a single relaxation time yields a half-decay lag measured
from the first lag (Section~\ref{sec:what-the-response-is}). (c)
Negative side: constant over anti-causal lags divided by the epoch mean
at the first lag. (d) $\Sodd/\Aeven$ at the first lag. Markers are
slightly shifted in $W$; the index fund’s 60-minute row, with six
epochs, is not plotted.}
\label{fig:scan}
\end{figure}

Table~\ref{tbl:robustness} in Appendix~\ref{app:robustness} compiles
checks that alter a single element of the analysis and rerun it, one row
per check; the amplitude across years and its dependence on volatility
level are treated in Section~\ref{sec:amplitude}. The rows appear in the
following order:
\begin{enumerate}
\item \emph{Daily cycle removed.} Prior to wing fitting each window’s
log variance is adjusted by subtracting the mean for its minute of day.
This mean is calculated over the other days of the epoch, in UTC for
cryptocurrencies and New York time for equities. $T$-even parts are
unchanged. $T$-odd parts for Bitcoin and Ether drop to roughly two
thirds of the $T$-even part while remaining significant
(Section~\ref{sec:directed-part}), scale-free amplitudes decrease by
about one tenth and the index fund’s mirror survives. The first design,
which also removed the mean trend per minute of day, failed the null
gate on the simulated market’s anti-causal lags and was abandoned
(Section~\ref{sec:null-gate}).
\item \emph{Linear channel.} Variance replaces its logarithm and is
normalized by the baseline of the two wings. The sign is unaltered in
every asset; for cryptocurrencies the amplitude is roughly twice as
large with the per-epoch difference $2.3$ to $3.8\sigma$ from zero;
equities agree within two standard errors.
\item \emph{First epoch removed.} Excluding 2020-01-20 to 2020-03-30,
which includes the March 2020 crash, no estimate changes by more than
0.6 of its error.
\item \emph{Falls versus rises.} In epochs where both wing decays are
resolved, 8 of Bitcoin’s 33, 10 of Ether’s and 10 of Solana’s 26, the
two relaxation times agree within approximately 20\%; no equity epoch
attains resolution.
\item \emph{Four-week epochs.} Using 84 epochs rather than 33 with no
other changes, the cryptocurrencies’ scale-free amplitudes shift by
roughly 15\%. This is within the errors of either run. For the index
fund the shorter epochs contain approximately 360 instants each and 29
of the 84 are excluded because more than 5\% of bootstrap replicas fail.
The fund’s 20-minute asymmetry is not comparable across the two
partitions.
\item \emph{The window length.} $W = 10$, 40 and 60 minutes as well as
20 for Bitcoin, Ether and the index fund, with the lag grid stepped by
$W$ and no other changes; the daily-cycle variants and the two Bitcoin
nulls were also assessed at each window (Fig.~\ref{fig:scan}). From 20
minutes upward the scale-free amplitude remains flat and the half-decay
lag stays between 86 and 120 minutes for Bitcoin and 90 and 100 minutes
for Ether; under the rule set before the runs, the value should lie
within the 20-minute interval, which three of the four cases meet.
Bitcoin at 60 minutes, 120 (104 to 153) versus an interval ending at
106, does not; this rise is examined in
Section~\ref{sec:what-the-response-is}. At $W = 10$ the response is
faster and weaker with a half-decay lag of 35 to 39 minutes: a component
within the first half hour that the 20-minute window averages over. The
negative side is resolved at some windows. At $W = 10$, where it is best
measured, it amounts to one sixth (Bitcoin) and one third (Ether) of the
first-lag value with opposite sign, and the joint anti-causal test
rejects zero; for Ether it is one half of the first-lag value at
$W = 40$ and comparable at 60. For Bitcoin it is roughly one third at
both windows, within errors of similar magnitude. A negative side equal
to one quarter of the first-lag value lies within approximately one
standard error of the constants measured at 20 minutes. Removing the
daily cycle does not eliminate it. The sign shuffle and simulated market
pass at the three new windows. The ratio $\Sodd/\Aeven$ therefore varies
with window: at the first lag it spans 0.4 to 0.9 at 10 and 20 minutes,
where the $T$-odd part is $2.6\sigma$ and $5.1\sigma$ from zero in
Bitcoin and $2.4\sigma$ and $4.0\sigma$ in Ether. The ratio is
unresolved at 40 and 60 minutes, where the $T$-odd error doubles and the
part lies within $1.7\sigma$ of zero. The quoted window produces the
highest ratio. For the index fund, $W = 10$ reproduces the 20-minute
mirror. At $W = 40$ the session has half as many windows; the negative
side maintains its value but loses precision. The first-lag asymmetry at
this window is nearly six times its 20-minute scale-free value (0.145
versus 0.026), so the two sides are no longer equal; at 60 minutes only
six epochs remain and no inference is possible. The rise is not a result
of sample size: for the fund at $W = 20$ with instants sampled every 40
minutes, the same sample as the 40-minute run, the 20-minute amplitude
is retained, $0.041 \pm 0.031$ versus $0.026 \pm 0.015$, while the
anti-causal test loses power and the anti-causal constant is unaltered.
Nor is it shared by the stocks: the three stocks at $W = 40$ shift by at
most 1.5 standard errors in either direction
(Table~\ref{tbl:robustness}). The rise of the index fund’s first-lag
asymmetry from 20 to 40 minutes has no parallel in cryptocurrencies or
stocks and remains unexplained (Section~\ref{sec:limits}).
\item \emph{Return spacing.} Realized variance calculated from
one-minute returns rather than two-second marks, with trend marks and
all downstream processing unaltered, for Bitcoin, Ether and the index
fund. Bid-ask bounce inflates two-second variance with a component that
does not respond to trend. This dilutes the slopes of the log channel
(Section~\ref{sec:limits}); one-minute returns contain little of this
effect, but only 20 of them fall within a window, so their log variance
is noisier and its mean is biased low — approximately 0.05 for Gaussian
returns, the expectation of the logarithm of a chi-square with 20
degrees of freedom divided by 20, and more for heavy-tailed returns. Per
epoch log variance at zero trend drops from two-second to one-minute
marks by $0.04 \pm 0.03$ for Bitcoin and $0.04 \pm 0.02$ for Ether,
consistent with that bias, and by $0.19 \pm 0.02$ for the index fund: a
noise contribution of approximately 0.14 in the logarithm of the index
fund’s two-second variance. None is detectable in the cryptocurrencies.
Scale-free amplitudes are larger at one minute by roughly one fifth per
epoch for Bitcoin and Ether and by $0.025 \pm 0.010$ for the index fund,
whose value is 0.026; $T$-odd parts change by about one standard error
or less, $T$-even parts grow and the index fund’s mirror survives. Marks
at 10 s, 30 s, 2 and 5 minutes for Bitcoin and the index fund
(Table~\ref{tbl:robustness}) place the one-minute values on a plateau
starting at 30 s and leave $T$-odd parts virtually unaltered. In Bitcoin
$\Sodd$ lies between $+0.029$ and $+0.034$ for all spacings from two
seconds to five minutes: spacing influences only the $T$-even part. The
two-second values reported here are therefore the lower ones. For
cryptocurrencies the difference cannot be accounted for by a noise
floor. We treat it as a systematic uncertainty in amplitude arising from
spacing choice, the second error listed in Table~\ref{tbl:main}.
\item \emph{The wing fit.} A quadratic term per wing for the seven
assets and a trend cut at 4 and 1.5 scales rather than 2.5 for Bitcoin,
Ether and the index fund. In the cryptocurrencies nothing changes: the
first-lag amplitude shifts by at most $0.004 \pm 0.011$ under the cuts
and by $+0.008 \pm 0.025$, $+0.018 \pm 0.026$ and $+0.012 \pm 0.022$ per
epoch with the quadratic term, whose inter-wing odd component — the
component that would bias a straight line — is $-0.002 \pm 0.006$ for
Bitcoin and $-0.004 \pm 0.006$ for Ether. Ether and Solana display
symmetric curvature that cancels. In the equities the fit matters: the
index fund’s amplitude is $0.026 \pm 0.015$ over 2.5 scales,
$0.019 \pm 0.013$ over 4, $0.110 \pm 0.026$ over 1.5 and $0.30 \pm 0.07$
with the quadratic term, whose odd component is $-0.071 \pm 0.017$; its
negative side diminishes from comparable to the positive side to one
quarter of it and then to zero. The three stocks behave similarly
(Section~\ref{sec:equities}).
\end{enumerate}

Four further checks address the index fund’s mirror and are presented
without a table. First, the clock-time test. In a simplified estimator
that fits a straight line through the 20-minute windows of a session
rather than binned wing fits, each window, its return and variance were
replaced by the window at the same clock time on a randomly selected day
within the same epoch. This retains the daily pattern while severing all
intraday links. The $T$-even part of the index fund declined from
$0.045 \pm 0.006$ to $0.001 \pm 0.003$ in those units. This is roughly
2\% of its original value. Second, trend persistence: the
autocorrelation of the index fund’s 20-minute returns is
$+0.009 \pm 0.008$ at 20 minutes and remains within $\pm 0.013$ at all
lags up to 120 minutes; limited to the most volatile third of prior
windows, where reversals after volatile moves would emerge, it is
$+0.012 \pm 0.010$ at 20 minutes and within $\pm 0.015$ at all lags;
limited to the quietest third it is $+0.020 \pm 0.008$ at 20 minutes and
within $\pm 0.02$ thereafter. Third, cleaning the tape
(Section~\ref{sec:data}) altered the equities’ anti-causal constants by
at most 0.7 standard errors. It reversed the sign of Apple’s asymmetry
over 20 to 180 minutes, from $-0.022 \pm 0.035$ to $+0.069 \pm 0.037$;
on the raw tape single off-market prints had dictated the variance of
9\% of the index fund’s 20-minute windows in the first quarter of 2020.
Fourth, the two channels of the index fund’s mirror, obtained from the
same window tables as the first check. The plain correlation of a
20-minute return with log variance 20 minutes later is $-0.11 \pm 0.01$
and stays at $-0.08$ two hours later. The correlation of log variance
with return 20 minutes later is $+0.013 \pm 0.005$ and is consistent
with zero at all longer lags; for Bitcoin the two correlations are
$-0.037 \pm 0.006$ and $+0.008 \pm 0.002$ at 20 minutes, and for Ether
$-0.050 \pm 0.006$ and $+0.010 \pm 0.002$. The return following a window
is more left-skewed after the quietest third of windows, with skewness
$-0.66 \pm 0.10$, than after the most volatile third, $-0.13 \pm 0.08$.
This compares with $-0.20$ at every level when windows are refilled from
other days at the same clock time. In Bitcoin and Ether the skewness of
the later return shows little dependence on prior variance, and in the
opposite direction: among returns within 2.5 standard deviations of
zero, the sample employed for wing fits, skewness is $+0.08 \pm 0.03$
after the quietest third and $-0.04 \pm 0.01$ after the most volatile
for Bitcoin, $+0.03 \pm 0.03$ and $-0.05 \pm 0.01$ for Ether. The index
fund in the same sample yields $-0.38 \pm 0.05$ and $-0.10 \pm 0.02$.
The interpretation of these values is addressed in
Section~\ref{sec:mirror}.

\section{Discussion}
\label{sec:discussion}

\subsection{What the cryptocurrency response is}
\label{sec:what-the-response-is}

The asymmetry of the three cryptocurrencies displays the profile of a
response function. It is small before the trend, peaks at the first lag
after it and decays over the ensuing hours. A variable that relaxes
toward its mean and is driven by a brief impulse behaves in this
fashion: no effect before the impulse and a one-sided decay thereafter
(Section~\ref{sec:quantities}). In time-reversal terms such a response
has equal $T$-odd and $T$-even parts. In Bitcoin and Ether the $T$-odd
part is comparable to the $T$-even part. This points to a one-sided
response in time within errors, with a $T$-odd part $4\sigma$ to
$5\sigma$ from zero. This sets it apart from the index fund
(Section~\ref{sec:mirror}). The scan over window lengths qualifies this
claim (Section~\ref{sec:robustness}). The negative side holds a
component of opposite sign, one sixth to one third of the first-lag
value at the 10-minute window, which the 20-minute window does not
resolve, and more so for Ether at 40 and 60 minutes; with this component
the $T$-odd part is 0.4 to 0.9 of the $T$-even part where it is measured
(Section~\ref{sec:directed-part}). The positive control in
Section~\ref{sec:nulls} was evaluated at the 20-minute window. It yields
the same picture for a purely one-sided response with Bitcoin’s
amplitude: a ratio of about 0.65 at the first lag and a negative side
one sixth to one quarter, stemming from the skewness imparted to the
trend by a one-sided response. At 20 minutes the response is therefore
consistent with being wholly one-sided in time. At that window the ratio
under one represents the statistic’s reading of trend skewness rather
than a backward-acting coupling. The control was not evaluated at the
other windows. Whether the same explanation accounts for the negative
side at 10 minutes and for Ether’s negative side at 40 and 60 minutes,
which amounts to one half of the first-lag value at 40 minutes and is as
large as it at 60, therefore remains unmeasured. Prior to the trend,
variance is slightly higher before a rise than before a fall, the sign
shown by the index fund’s negative side. The correlation statistic from
Section~\ref{sec:robustness} conveys the same finding for Bitcoin at 20
minutes: the correlation of log variance with the later return is
$+0.008 \pm 0.002$, about one fifth of the correlation of return with
later log variance, $-0.037 \pm 0.006$, and of opposite sign. This
component does not share the index fund’s origin, namely the shape of
the later return (Section~\ref{sec:mirror}).

What the data establish is temporal order, with the trend preceding the
excess volatility. We therefore state that the asymmetry follows the
trend and suspect the trend might cause it. The data reveal no more than
this. A common driver, such as news that moves price first and
volatility later, would generate the same profile. Negative lags
demonstrate that a coupling without temporal direction does not explain
all of it, but they do not rule out a common driver.

The magnitude is modest. For a trend of typical size, variance following
a fall is about 7\% higher than following a rise. This figure is the
same for Bitcoin and Ether within errors. In both assets the effect is
durable: positive in every year from 2020 to 2026, with the two halves
of the period in agreement. In Solana the effect is about one third as
large and arises in the second half of its record
(Section~\ref{sec:amplitude}). For Bitcoin and Ether the coupling is an
enduring property rather than an episode. We interpret only the
asymmetry. The symmetric component, the increase in volatility after a
move of either sign, is much larger but chiefly reflects that large
moves occur in volatile periods (Section~\ref{sec:response-function}). A
fraction of the $T$-odd signal is linked to time of day: removing the
daily volatility cycle lowers Bitcoin’s $T$-odd part by about one
quarter and Ether’s by about one seventh (Section~\ref{sec:robustness}).
Whether this fraction reflects a response absorbed by the removal or a
coincidence between trend sign and hour cannot be settled; we regard it
as a systematic uncertainty in the $T$-odd part.

Two features of the decay warrant comment. The response halves by a lag
of about one and a half hours for windows of 20 to 60 minutes, amounting
to 86 to 120 minutes; a 10-minute window resolves a faster component
within the first half hour (Section~\ref{sec:robustness}). The
one-and-a-half-hour scale is thus the slower of at least two. The rise
in half-decay lag with window length from 20 to 60 minutes stems from
the shift in the first lag rather than from a change in decay. The
half-decay lag is measured from the first lag, $\ell = W$. Under the
assumed form the response at every lag $\ell \ge W$ is the same
exponential for each window up to a constant; with no floor the
half-decay lag measured from the first lag is $\tau \ln 2$ regardless of
$W$, whereas a pure power law with no scale and no floor would make it
proportional to $W$. The scan agrees with the first case: the half-decay
lag minus $W$ is 69, 46 and 60 minutes for Bitcoin and 70, 60 and 36
minutes for Ether at $W = 20$, 40 and 60, constant within intervals and,
for Bitcoin, consistent with 66 minutes implied by $\tau = 95$ minutes
(dotted line in Fig.~\ref{fig:scan}b). In contrast the ratio of
half-decay lag to window falls by half. The only value violating the
rule in Section~\ref{sec:robustness}, Bitcoin at 60 minutes, results
from this shift. At $W = 10$ the shifted lag is 29 and 25 minutes, once
more the faster component; four windows cannot rule out a scale-free
kernel over a broader range, nor do they test the power law with a floor
from Appendix~\ref{app:shape}. For US stocks an intraday leverage kernel
decaying in about an hour has been mentioned in passing from five-minute
returns \cite{blanc2014}; whether the two time scales are related cannot
be ascertained. In Ether, and less distinctly in Solana, the response
then levels off within our ten-hour window at a floor of about one third
of its 20-minute value. The floor captures modes slower than the window;
a mode slower than a day is constant within it. It cannot be permanent:
an enduring imprint of every trend would connect log variance to log
price. We expect it to relax on a longer scale, which was not examined.
The shape within the window is assumed rather than measured
(Section~\ref{sec:limits}). At two-minute sampling the return–volatility
cross-correlation of Bitcoin shows the same sign and a slow decay fitted
by a power law \cite{takaishi2020}; our data cannot discriminate between
a single relaxation time with a floor and a power law with a floor
(Section~\ref{sec:time-scale}).

\subsection{Why daily studies saw little}
\label{sec:daily-studies}

Most of the response we measure unfolds within hours
(Section~\ref{sec:time-scale}): in Bitcoin and Ether it halves by a lag
of about one and a half hours, and under the fitted form after four to
five hours it attains its floor, consistent with zero for Bitcoin and
about one third of the 20-minute value for Ether. A daily estimator
registers little of this. A threshold GARCH connects the variance of one
day to the sign of the previous day’s return, as does any daily model.
Its lag is roughly one day. At that lag the decaying component is
diluted to a small remnant because only the final hours of the preceding
day fall within a few relaxation times of the initial windows of the
next. What remains is the floor, if it persists for a day. The model
weighs it against the variance of a full day. Under the assumed form and
linear response the mean asymmetry over pairs of 20-minute windows in
successive days is 5\% of the 20-minute value for Bitcoin and 35\% for
Ether if the floor persists for a day, and under 1\% for both if it does
not; this is an extrapolation from a form fitted within ten hours, not a
direct measurement. The null result of Baur and Dimpfl for Bitcoin and
Ether from daily data up to 2018 \cite{baur2018} is consistent with a
response of this time scale. Their quantile measure, which uncovers the
stock-like pattern for Bitcoin, has the same sign as ours. The periods
do not overlap because our data begin in 2020. What we add is the
component of the response that daily data cannot resolve. Within the
first ten hours this component makes up most of it. Among the mixed
daily findings on cryptocurrencies (Section~\ref{sec:introduction}),
null results may stem from the same limitation; a reversed sign cannot.
Horizon also matters in intraday studies: in realized semivariances of
Bitcoin the asymmetry is reversed at one day and stock-like at seven
days \cite{jha2020}, and the inverted term of a threshold GARCH weakens
over the years \cite{takaishi2021}.

For equities both daily and intraday pictures have been reported. At
daily lags the leverage correlation of stock indices is pronounced,
one-sided in time and decays over about ten days \cite{bouchaud2001}. In
S\&P~500 futures at five-minute resolution, Bollerslev, Litvinova and
Tauchen found the correlation between a return and the later absolute
returns to be negative from lag zero and decaying over at least five
days, while correlations in the opposite direction hovered around zero
\cite{bollerslev2006}. In a similar statistic our index fund agrees with
them in both directions: the correlation of a 20-minute return with the
ensuing log variance is $-0.11$ and stays $-0.08$ two hours later, while
the correlation of log variance with the later return is
$+0.013 \pm 0.005$ at 20 minutes and consistent with zero thereafter
(Section~\ref{sec:robustness}). The mirror of the wing asymmetry is a
distinct statistic, into which the shape of the later return enters;
Section~\ref{sec:mirror} discusses it.

\subsection{The index fund's mirror}
\label{sec:mirror}

The index fund’s $T$-even part is large, as significant at 20 minutes as
Bitcoin’s. Its asymmetry is not a one-sided response of the sort seen in
cryptocurrencies. At positive lags the asymmetry is positive and
significant, as expected for a leverage effect, but at negative lags it
is equally large with opposite sign. Its $T$-odd part is consistent with
zero at every lag up to two hours, where both sides of a lag lie within
the session. Within this horizon we detect no arrow of time in the wing
slopes. In plain terms: a rise is preceded by higher volatility and
followed by lower volatility than a fall of the same magnitude. The
magnitude of this tilt is steady from 20 to 120 minutes within errors.
Beyond two hours the index fund’s asymmetry climbs to its largest
values, but the negative side cannot be measured there. The remaining
instants pair a morning trend with an afternoon volatility window. The
nature of the coupling in that regime remains unknown.

We sought an artifact that could create the mirror and found none
(Section~\ref{sec:robustness}). It remains after cleaning the tape,
which shifted the equities’ anti-causal constants by less than one
standard error. It is not a calendar effect: data that retain the
complete trading-day pattern while severing all intraday links yield 2\%
of the $T$-even part. Removing the intraday volatility profile leaves it
unchanged. It is not trend persistence: the index fund’s 20-minute
returns are uncorrelated with later returns up to two hours. Nor is it a
reversal that follows volatile moves only, which simple autocorrelation
would overlook: within the most volatile third of windows later returns
are as uncorrelated as within the quietest third
(Section~\ref{sec:robustness}). Moreover the minute shuffle eliminates
the mirror (Table~\ref{tbl:nulls}): once the net direction of each
minute is randomized while maintaining intra-minute paths, session
structure and overnight gaps, it vanishes. The mirror therefore stems
from the directions of moves within the trading day and their connection
to surrounding volatility.

The two sides of the mirror have distinct origins. With the window
tables from the clock-time test we checked whether the negative-lag side
represents a mean effect — expected return growing with prior variance,
the direction of volatility feedback — or a shape effect, where the
distribution of the later return varies with prior variance
(Section~\ref{sec:robustness}). It is a shape effect. The correlation of
log variance with the later return is small, $+0.013 \pm 0.005$ at 20
minutes and consistent with zero thereafter. The mean of that return is
independent of prior variance within errors; its skewness is not. The
return is markedly left-skewed following quiet windows and almost
symmetric following volatile ones. A dependence of this form has a
precedent at daily lags: in Qiu et al.’s words, “after the volatility
[...] is continuously stable for a period of time, the price drops”
\cite{qiu2006}. Such a law yields more small rises and more large falls
after quiet windows than after volatile ones. The wing fit at a negative
lag interprets this as higher variance before a rise. The positive-lag
side differs. There the correlation of return with ensuing log variance
is $-0.11$ and lasts for two hours: later volatility responds to the
earlier move in the leverage direction, as Bollerslev, Litvinova and
Tauchen reported in five-minute S\&P~500 futures data
\cite{bollerslev2006}. Most of this response is linked to the sign of
the move rather than its magnitude. In the simplified estimator from the
clock-time test the two wings differ in intercept at zero trend by about
0.12 in log variance at 20 minutes, whereas the wing slopes — our
statistic — account for the smaller share. The wing asymmetry cannot
disentangle the two origins and registers them as a near-equal mirror.
The index fund is therefore neither a one-sided response like that of
cryptocurrencies nor a coupling without temporal direction: in
correlation its arrow of time is as evident as Bitcoin’s. No mechanism
is claimed for either side. The negative side resolved by the scan in
cryptocurrencies does not share the index fund’s origin: there the shape
of the later return varies little with prior variance, and the two
channels display a small positive reverse correlation,
$+0.008 \pm 0.002$ for Bitcoin and $+0.010 \pm 0.002$ for Ether, with
expected return growing slightly with prior variance
(Section~\ref{sec:robustness}). The dependence of the fund’s asymmetry
on the trend range used for wing fitting (Section~\ref{sec:equities})
points the same way: a response to the sign of the move is concentrated
near zero trend, and a slope fitted over a wide trend range averages it
away.

The methodological point we would highlight is this. The jump in
asymmetry at lag zero does not differentiate between the two markets.
Negative lags do, by the magnitude of the negative side relative to the
positive side, which is comparable across the band for the index fund
and one sixth to one third of the first-lag value for cryptocurrencies
at the finest window (Section~\ref{sec:robustness}). They also call for
more than one statistic: wing asymmetry alone would have taken the index
fund’s mirror for time symmetry, and only correlation and the shape of
the later return uncovered the nature of its two sides. How the index
fund’s intraday response relates to the one-sided leverage effect of
indices at daily lags \cite{bouchaud2001} lies beyond intraday data
(Section~\ref{sec:equities}).

\subsection{Why the two markets may differ}
\label{sec:why-differ}

We do not know why the two markets differ. The following is untested
speculation. The difference is one of degree: cryptocurrencies display a
smaller component of the same sign at negative lags
(Section~\ref{sec:robustness}). The speculation pertains to the
one-sided response overlaid on it. A cryptocurrency lacks a fundamental
anchor. There is no firm, balance sheet or cash flow behind it; its
price is determined by speculative demand and trader behavior. This
alone fixes no sign. It implies that activity, and hence volatility, may
grow after a large move in either direction, which is the symmetric
component. The asymmetry calls for an explanation of why a fall
increases volatility more than a rise. The explanation that gave the
effect its name — firm debt \cite{black1976} — cannot apply. Volatility
feedback acts in the opposite temporal direction, with volatility
preceding price. The effect it would create, a return increasing with
prior variance, is small in Bitcoin, about one fifth of the forward
correlation (Section~\ref{sec:what-the-response-is}). A natural
candidate is trader leverage. Much cryptocurrency trading takes place in
perpetual futures, whose volume is two to three times that of spot
markets, with leverage up to one hundred times \cite{he2022,soska2021}.
When price falls, leveraged long positions are liquidated by the
exchange, a few percent of open interest on a typical day and more on
the long side than the short \cite{soska2021,cheng2021}, adding selling
pressure and volatility. This would yield the observed sign if leverage
is long-biased, and would operate within the day if liquidations unfold
over hours. The data used here are spot trades on a single exchange,
whereas liquidations occur on derivatives venues and propagate to spot
prices via arbitrage. A threshold mechanism of this type would also bend
the down wing, but quadratic fits in Table~\ref{tbl:robustness} reveal
no wing-specific curvature: the odd component of curvature is
$-0.002 \pm 0.006$ for Bitcoin and $-0.004 \pm 0.006$ for Ether. Data on
forced liquidations could test the mechanism directly.

The index fund differs in every aspect of this picture. It has a
valuation; a session with an open, a close and an overnight gap; a
pronounced intraday volatility profile; hedging flows that cluster at
the close \cite{baltussen2021}; and a large options market. We cannot
determine which of these gives rise to the two sides of its mirror — the
response to the sign of a move after it and the dependence of a move’s
shape on prior variance. One thing can be ruled out: the difference is
not calendar-driven, as the calendar alone accounts for nearly none of
the mirror (Section~\ref{sec:robustness}). It lies in intraday dynamics.

\subsection{Limits}
\label{sec:limits}

Null data are available for Bitcoin and the index fund only, and
cryptocurrency data are drawn from a single exchange. The observable is
the logarithm of a realized variance computed from two-second marks.
Microstructure noise adds to variance, particularly in thinner tapes. A
noise floor compresses log variance, shrinking every slope and asymmetry
toward zero by an asset-dependent amount. The seventh check in
Section~\ref{sec:robustness} quantifies this: replacing two-second marks
with one-minute returns raises scale-free amplitudes by about one fifth
for Bitcoin and Ether and by about 0.025 for the index fund, whose value
is 0.026, with $T$-odd parts unaltered within errors. The two-second
values reported here are the smaller ones. The ratio between
cryptocurrencies and the index fund (Section~\ref{sec:equities}) is most
sensitive to this choice: 2.7 at two seconds versus 2.1 at one minute.
Solana’s record is shorter and its evidence weaker; its $T$-odd part is
not significant by itself. The window length is $W = 20$ minutes
throughout, with a scan over 10, 40 and 60 minutes for Bitcoin, Ether
and the index fund (Section~\ref{sec:robustness}, Fig.~\ref{fig:scan}).
Amplitude is steady from 20 minutes upward and the half-decay lag lies
between 86 and 120 minutes; the first measurable lag depends on the
window. The faster component resolved by a 10-minute window has a time
scale that was not explored. The index fund’s mirror is established at
10 and 20 minutes; at 40 minutes its negative side is unresolved and its
positive side is much larger than at 20. This effect is not due to the
smaller sample of that window, has no equivalent in cryptocurrencies or
stocks and remains unexplained. The fund’s first-lag asymmetry and
$T$-odd part are quoted for the straight line over 2.5 trend scales. The
mirror is a feature of that fit: over inner trends and with a quadratic
term the fund’s positive side rises by a factor of four to twelve while
its negative side decreases, whereas Bitcoin’s and Ether’s values do not
change (Section~\ref{sec:robustness}). Compared with the positive
control the fund’s figures over inner trends do not distinguish between
a one-sided response and a mirror (Section~\ref{sec:equities}). The
relaxation time is based on a single method. The second method specified
in advance returns 248 minutes for Bitcoin versus 95 and fails its own
checks (Appendix~\ref{app:shape}). The half-decay lag, the time scale
reported here, is biased downward for a one-sided response, by one tenth
to one sixth in the positive control, whereas $\tau$ with a floor is not
(Sections~\ref{sec:time-scale} and~\ref{sec:nulls}). The decay shape is
assumed. A single relaxation time with a floor describes the data, as
does a power law with a floor. Our test favored the exponential even on
data without coupling (Section~\ref{sec:nulls}). We therefore offer the
form as a model consistent with the data, not as an established result.
The tail of the response was not investigated. At long lags ensuing
trend impulses fall between the two windows, so the response no longer
represents the decay of a single impulse. For equities the relaxation
time cannot be measured from intraday data because the closed market
leaves lags from about five and a half to eighteen hours empty.

\section{Conclusions}
\label{sec:conclusions}

We measured how market volatility responds to a preceding price trend
lag by lag at minute resolution in Bitcoin, Ether and Solana on a single
exchange and in three large US stocks and the S\&P~500 exchange-traded
fund from 2020 to 2026. We did this for both positive and negative lags
so that a response to the trend can be distinguished from a coupling
without temporal direction.

In the cryptocurrencies the asymmetry — excess volatility after a fall
relative to a rise of the same magnitude — is positive and follows the
trend: highest at the first measurable lag and small at negative lags,
where a scan over window lengths resolves a component of opposite sign
that has not yet been disentangled from what the statistic itself
creates. In Bitcoin and Ether it halves by a lag of about one and a half
hours, is identical within errors for both and positive in every year
since 2020; Solana shows about one third of this effect. The part that
changes sign under time reversal, which alone carries an arrow of time,
is 0.4 to 0.9 of the part that does not at the first lag of the 10- and
20-minute windows where it is detected, and is unresolved at 40 and 60
minutes; at 20 minutes a purely one-sided response of the same amplitude
embedded in a simulated market produces the same ratio within errors. At
that window the response is therefore consistent with being wholly
one-sided in time. We note that the asymmetry follows the trend and
suspect the trend might cause it; the data show only that the trend
precedes it. Most of the response unfolds within the first hours, a
feature that daily data cannot resolve.

The equities, measured with the same instrument, behave differently.
None of the four shows a detectable $T$-odd part at 20 minutes in the
straight-line fit over 2.5 trend scales, and the index fund shows none
up to two hours. In that fit the index fund’s asymmetry is mirrored in
lag, of equal size at negative and positive lags with opposite sign. The
two sides stem from distinct origins: after a move, volatility
responding to the sign of the move in the leverage direction; before a
move, the shape of the move varying with prior variance. Over inner
trends the fund’s positive side is four times larger and its negative
side is one quarter of it, which does not distinguish between a
one-sided response and a mirror; the fund responds to the sign of a move
more than to its magnitude, an effect poorly captured by a slope fitted
over a wide trend range, whereas the cryptocurrencies’ asymmetry is
independent of the shape of the fit, and Bitcoin’s and Ether’s of its
range. Our claim holds within the trading day, for windows of 10 and 20
minutes and for that fit; at daily lags the leverage effect of indices
is well established \cite{bouchaud2001}. Why the two markets differ is
still unknown (Section~\ref{sec:why-differ}).

Two methodological points apply beyond this study. Without negative lags
the index fund’s wing asymmetry would have been confused with a
one-sided response like that of cryptocurrencies. Because neighboring
lags are almost redundant, all errors must be obtained from independent
data blocks, here ten-week epochs; the full analysis yields zero on data
without coupling. The decay shape is assumed rather than measured. The
tail beyond ten hours was not investigated. A lag axis that bypasses
closed markets, data on forced liquidations, and a second
relaxation-time method that passes its own checks are natural next
steps.

\appendix
\numberwithin{table}{section}
\numberwithin{figure}{section}
\numberwithin{equation}{section}

\section{The error model}
\label{app:error-model}

\paragraph{Within an epoch}
The day-block bootstrap in Section~\ref{sec:errors} resamples epoch days
with replacement in blocks of five consecutive days and recomputes all
downstream quantities for each replica. Hence the error of a derived
quantity is the dispersion of that quantity itself rather than a
delta-method propagation from inputs. The same days contribute to every
bin and lag in a replica. Day-level commonalities thus move together and
the replicas retain correlations across bins and lags. Draws are seeded
from the configuration, block and epoch. Reruns are therefore identical.
Wing-fit weights are the bins’ naive standard errors. Each is the bin
standard deviation divided by the square root of sample size; the bars
in Fig.~\ref{fig:measurement} instead display each bin’s dispersion
across day-block replicas, which accounts for within-day sample
correlation and is larger. Measured against naive weights, the straight
lines of the primary analysis have a median $\chi^2$ per degree of
freedom of 1.07 across the 12,404 wing fits for all assets, lags and
epochs (1.13 for cryptocurrencies, 0.93 for equities with one to four
degrees of freedom), and 13\% of fits exceed 2, a check against errors
known to be too small; against replica dispersions the fit shown in
Fig.~\ref{fig:measurement} has 1.3 and 0.5 for its two wings where naive
weights yield 2.8 and 1.5. A check against the full bin covariance is
infeasible: a twelve-bin covariance cannot be inverted from fourteen day
blocks. The bootstrap error of a slope is twice the fit’s own for
cryptocurrencies and 1.1 to 1.2 times for equities, as medians: bin
samples are correlated and bins share days. Hence no per-epoch error is
taken from the fit. The band in Fig.~\ref{fig:measurement} is the fit’s
own scaled by the ratio of slope errors, 2.2 and 1.5 for its two wings.
Four flags accompany every per-epoch value. Flagged values are excluded
from pooling: \emph{not fit}, a wing with fewer than three usable bins;
\emph{without replicas}, a value whose sole error is the delta-method
error of the fit; \emph{censored}, more than 5\% of the 200 replicas
failed, so the dispersion of survivors would narrow as the fit
deteriorates; and for the relative asymmetry of the linear channel,
which is a ratio, \emph{withheld}, when any replica’s baseline crosses
zero or the baseline is less than five standard errors from zero.

\paragraph{What enters each pooled value}
Each row entering a pool is counted once by the first rule it violates,
in the order overlap, not fit, without replicas, censored, withheld; an
epoch with no row at a lag is counted as absent. For cryptocurrencies
every epoch is included at every lag except the overlapping lag. For
equities lags beyond the session contain no instants; from 240 to 320
minutes the session provides few instants per day and up to 31 of the 33
epochs are censored at a given lag. Pooled curves for equities terminate
at 260 minutes (280 for the index fund). In the log channel no asset has
values without replicas or withheld.

\paragraph{Across epochs}
The pooled value at each lag is the inverse-variance weighted mean of
the epoch values, and its error is the weighted error multiplied by
$S = \max(1, \sqrt{\chi^2/\mathrm{dof}})$. Here $\chi^2$ measures epoch
scatter about the mean in units of their errors; $S$ is never below one
because deflation would conceal the common mode the rule is intended to
capture. In addition an epoch-block bootstrap resamples epochs 400 times
in blocks of two consecutive epochs and reports the dispersion of the
recomputed mean; it uses only epoch values, not their errors. The pool
also reports the unweighted mean and median, the correlation between
epoch values and their errors, and the largest weight. It flags a lag
where errors track values, the signature of weighting biased toward
quiet epochs. In the log channel the flag is raised at no lag for
Bitcoin and Apple, and at one to three isolated lags for each other
asset. In this channel it is never raised at 20 minutes. In the linear
channel the flag is raised more frequently, at 20 minutes for three
assets. This is one reason the log channel is primary.

\paragraph{The joint tests}
The statistic $Q$ from Section~\ref{sec:errors} is the sum over the lags
in a region of the squared pooled value in units of its error. Its null
distribution is obtained by resampling. Per-epoch values are demeaned at
each lag so the null hypothesis holds exactly in the resampled data;
epochs are resampled in circular blocks of two consecutive epochs
jointly across lags, retaining lag correlations; pooled values and $Q$
are recomputed for each of 400 replicas; and $p = (k+1)/(B+1)$, where
$k$ of the $B$ replicas attain the observed $Q$. The smallest achievable
$p$ is $1/401$, an upper bound. The effective number of independent lags
is $\operatorname{tr}(C)^2/\operatorname{tr}(C^2)$ for the correlation
matrix $C$ of per-epoch residuals across lags in the region; it equals
the number of lags when independent and one when perfectly correlated
(Table~\ref{tbl:effective-lags}).

\begin{table}[htp]
\centering
\caption{The effective number of independent lags,
$\operatorname{tr}(C)^2/\operatorname{tr}(C^2)$, for the pooled
asymmetry curves in the log channel, by region. The equities’ causal
region terminates within the session.}
\label{tbl:effective-lags}
\begin{tabular}{@{}lrrrr@{}}
\toprule
asset & causal lags & effectively independent & anti-causal lags & effectively independent \\
\midrule
BTC  & 30 & 3.5 & 6 & 1.7 \\
ETH  & 30 & 2.9 & 6 & 1.8 \\
SOL  & 30 & 2.9 & 6 & 1.5 \\
AAPL & 13 & 5.0 & 6 & 2.6 \\
TSLA & 13 & 3.0 & 6 & 2.2 \\
MSFT & 13 & 5.2 & 6 & 2.3 \\
SPY  & 14 & 4.7 & 6 & 2.2 \\
\bottomrule
\end{tabular}
\end{table}

\paragraph{Fitted quantities}
Each quantity fitted to a pooled curve obtains its interval from a
bootstrap of the full analysis over epochs: epochs are resampled, the
curve is re-pooled, the quantity is recomputed and the interval is read
from the percentiles of the replicas (1,500 for the decay in
Section~\ref{sec:quantities}, 4,000 for the ratio $\Sodd/\Aeven$). In a
previous geometry with 25 lag points, correlation between neighboring
lags across epochs was 0.55 to 0.77. In that geometry the error of a
fitted relaxation time derived from lag-by-lag covariance under the
assumption of independent lags was smaller than the epoch-bootstrap
error by factors of 64 to 859. The symptom was a $\chi^2$ per degree of
freedom of 0.2 to 0.6 with $p$ close to one: a deflated $\chi^2$, not a
good fit.

\section{The relaxation time and the shape}
\label{app:shape}

\paragraph{The fit of the assumed form}
The pooled curve is the unweighted mean of per-epoch asymmetry at each
lag with its standard error across epochs (Section~\ref{sec:errors}).
The form $b + a\,e^{-\ell/\tau}$ is fitted to it over the 30 causal
lags: for each $\tau$ on a grid of 1,200 logarithmically spaced values
from 1 to 6,000 minutes, $b$ and $a$ are the weighted least-squares
solution using per-lag standard errors as weights. The $\tau$ with the
smallest weighted residual sum is chosen. The residual sum is not a
$\chi^2$ because lags are correlated. No inference is drawn from it.
Intervals for $\tau$, $b$ and the half-decay lag are the 16th to 84th
percentiles across 1,500 replicas of the full analysis. In each replica
epochs are resampled, the curve is re-pooled and the fit is repeated;
the band in Fig.~\ref{fig:response} is the standard deviation of the
fitted curve across the same replicas. Table~\ref{tbl:time-scale} lists
the values. With approximately three effectively independent lags
(Table~\ref{tbl:effective-lags}), three parameters is the capacity of
the window; two relaxation times, a stretched exponential or a crossover
to a power law are not identifiable here and were not attempted. The
upper limit of the $\tau$ grid, 6,000 minutes, is ten times the window:
a replica whose $\tau$ reaches it exhibits no resolved decay. No Bitcoin
replica reaches it, one in 1,500 for Ether and 9\% for Solana.

\paragraph{The method specified in advance}
Before the analysis a second estimator for $\tau$ was defined, a
likelihood pooled over epochs: per epoch the asymmetry curve over causal
lags and its 200 day-block bootstrap replicas provide a lag-by-lag
covariance, the form is fitted by generalized least squares on a grid of
400 $\tau$ values from 1 to 600 minutes and per-epoch $\chi^2$ curves
are summed; $\tau$ is the minimum of the sum and its interval is the
rise by one. With 14 resampling units against 30 lags the replica
covariance has rank at most 14 and cannot be inverted. Consequently the
estimator weights each lag by its own replica variance, ignores
inter-lag correlations, and the sum is a quasi-likelihood; therefore
three checks were fixed in advance (the rise-by-one interval must match
the dispersion of $\tau$ over replica-by-replica poolings, a held-out
replica must yield $\chi^2$ per degree of freedom within a factor of
three of one, and the result must be stable when the grid is broadened
to 0.5 to 1,200 minutes) and a control (on the common response, which
carries no such decay, it must find no confident $\tau$). It resolves
$\tau$ for Bitcoin only, at 248 minutes (188 to 350) versus 95 from the
pooled-curve fit, with the first interval 1.26 times the replica
dispersion where noise permits $1 \pm 0.06$, a calibration of 0.41 and
eight of the 33 epochs on the grid edge; for the other assets it does
not resolve. The broadened grid alters every verdict except Ether’s, and
the control fails, giving 205 minutes for Bitcoin’s common response.
Where it does resolve, $\tau$ lies at 0.13 to 0.52 of the 600-minute
window, the signature of a fit that has not disentangled decay from the
window. The checks were not satisfied and the estimator is not used;
$\tau$ therefore appears in the article as a parameter of an assumed
model and the half-decay lag as the time scale.

\paragraph{The shape comparison and its gate}
To test whether the data favor one relaxation time over a power law, a
second three-parameter form was fitted to the same pooled curve,
$b + a\,(\ell / 20~\text{min})^{-\alpha}$ with $\alpha$ on a logarithmic
grid of 400 values from 0.02 to 5. Both forms were refitted in every
bootstrap replica and the fraction $f$ of replicas in which the
exponential has the smaller weighted residual sum was counted; a shape
was to be deemed preferred at $f \ge 0.9$ or $f \le 0.1$, a threshold
set before the test. By the rule in Section~\ref{sec:null-gate} the test
had to leave both shapes unpreferred on every null. It did not: on the
six Bitcoin nulls the exponential wins in 95 to 100\% of replicas
(Table~\ref{tbl:shape}). This is because with a free floor and $\tau$ up
to 6,000 minutes the exponential family includes a straight line in
$\ell$ whereas the long-scale limit of the power law is a line in
$\ln \ell$, and a noise curve with linear drift across the window, which
the nulls display, is fitted better by the former. The preference is
determined by the behavior of the families at their grid edges, not by
the shape of a decay. The test is therefore inconclusive; its values on
the data, 0.85, 0.42 and 0.87 for Bitcoin, Ether and Solana, are not
interpreted. The failure is not evidence against the exponential: on the
data the two fitted curves lie within a fraction of a standard error of
each other at every lag.

\begin{table}[htp]
\centering
\caption{Shape comparison. For each curve the fraction $f$ of the 1,500
epoch-bootstrap replicas in which the exponential with a floor has the
smaller weighted residual sum than the power law with a floor, the two
fitted forms and the residual sums over the 30 causal lags weighted by
per-lag standard errors (not a $\chi^2$ because lags are correlated). A
shape would have been deemed preferred at $f \ge 0.9$ or $f \le 0.1$;
the six Bitcoin nulls fail the gate. The test is therefore not
interpreted. “Edge” indicates a scale at the end of its grid; the
interval is across replicas, whose median may differ from the
full-sample fit.}
\label{tbl:shape}
\scriptsize
\begin{tabular}{@{}lrllr@{}}
\toprule
data & $f$ & exponential: $\tau$ [min], $b$ & power law: $\alpha$ (16th to 84th), $b$ & \makecell[r]{residual sums,\\exponential / power law} \\
\midrule
Bitcoin                   & 0.85 & 95, $+0.008$        & 0.39 (0.15 to 0.65), $-0.071$ & 13.0 / 14.5 \\
Ether                     & 0.42 & 53, $+0.044$        & 0.65 (0.20 to 1.31), $+0.028$ & 23.3 / 21.7 \\
Solana                    & 0.87 & 62, $+0.016$        & 0.82 (0.13 to 1.40), $+0.013$ &  7.3 /  8.0 \\
\midrule
simulated market, seed 1  & 0.95 & 359, $-0.006$       & 0.02 (edge)                   & 13.2 / 14.5 \\
simulated market, seed 2  & 0.99 & 520, $+0.012$       & 0.02 (edge)                   &  5.9 /  7.5 \\
simulated market, seed 3  & 0.97 & 6,000 (edge), $+0.028$ & 3.5 (0.02 to 3.2), $+0.003$ &  5.7 /  5.8 \\
sign shuffle, seed 1      & 1.00 & 6,000 (edge), $+0.67$ & 0.02 (edge)                 & 21.7 / 30.4 \\
sign shuffle, seed 2      & 0.99 & 44, $-0.045$        & 5.0 (edge)                    & 10.7 / 10.7 \\
sign shuffle, seed 3      & 1.00 & 6,000 (edge), $+0.53$ & 0.02 (edge)                 & 21.6 / 25.3 \\
\bottomrule
\end{tabular}
\end{table}

\section{Robustness}
\label{app:robustness}

Table~\ref{tbl:robustness} compiles the checks from
Section~\ref{sec:robustness}. Each varies a single element of the
analysis and repeats it; the text discusses each row.

\begingroup
\scriptsize
\setlength{\tabcolsep}{4pt}
\setlength{\LTcapwidth}{\textwidth}
\newlength{\robustnessunit}
\setlength{\robustnessunit}{\dimexpr(\linewidth-8\tabcolsep)/148\relax}
\begin{longtable}{@{}>{\raggedright\arraybackslash}p{19\robustnessunit}>{\raggedright\arraybackslash}p{33\robustnessunit}>{\raggedright\arraybackslash}p{32\robustnessunit}>{\raggedright\arraybackslash}p{32\robustnessunit}>{\raggedright\arraybackslash}p{32\robustnessunit}@{}}
\caption{Robustness. Each row varies one element of the analysis and
repeats it; values are reported at 20 minutes unless otherwise stated,
in the units of their respective tables. $\Aeven$ and $\Sodd$ are the
$T$-even and $T$-odd parts (Table~\ref{tbl:time-reversal}), $\Delta_s$
is the scale-free amplitude (Table~\ref{tbl:main}), the band is the mean
asymmetry over 20 to 180 minutes and $\tau_\downarrow/\tau_\uparrow$ is
the ratio of relaxation times after falls and after rises, with its 68\%
interval, for epochs where both are resolved. In the window-length row
the first-lag value is the epoch mean $\Aeven + \Sodd$ and the
anti-causal constant is the epoch mean over anti-causal lags; in the
return-spacing row “level” denotes the change in log variance at zero
trend per epoch, and the per-epoch change in $\Delta_s$ is shown in
brackets.}\label{tbl:robustness}\\
\toprule
check & what changes & Bitcoin & Ether & the index fund \\
\midrule
\endfirsthead
\caption[]{Robustness (continued).}\\
\toprule
check & what changes & Bitcoin & Ether & the index fund \\
\midrule
\endhead
\bottomrule
\endfoot
daily cycle removed & every window's log variance loses the mean at its minute of the day, over the other days of the epoch &
  $\Aeven$ $+0.038 \to +0.038$; $\Sodd$ $+0.032 \to +0.025$ ($3.8\sigma$); $\Delta_s$ $0.071 \to 0.064$; anti-causal $p$ $0.46 \to 0.12$ &
  $\Aeven$ $+0.037 \to +0.037$; $\Sodd$ $+0.026 \to +0.023$ ($4.1\sigma$); $\Delta_s$ $0.066 \to 0.058$; anti-causal $p$ $0.58 \to 0.05$ &
  $\Aeven$ $+0.051 \to +0.050$; $\Sodd$ $-0.025 \to -0.012$; anti-causal $p$ $0.005 \to {<}\,0.0025$: the mirror stays \\
\addlinespace
the linear channel & the variance in place of its logarithm, divided by the wings' baseline &
  $\Delta_s$ $0.071 \pm 0.008 \to 0.130 \pm 0.026$; the difference per epoch $-2.3\sigma$ &
  $0.066 \pm 0.008 \to 0.141 \pm 0.022$; $-3.8\sigma$ &
  $0.026 \pm 0.015 \to 0.025 \pm 0.041$; $-0.4\sigma$ \\
\addlinespace
first epoch removed & the epoch 2020-01-20 to 2020-03-30 dropped &
  $\Delta_s$ $0.071 \to 0.071$; band $+0.093 \to +0.092$; $\Sodd$ $+0.032 \to +0.032$; $\tau$ $95 \to 96$ min &
  $\Delta_s$ $0.066 \to 0.062$; band $+0.070 \to +0.069$; $\Sodd$ $+0.026 \to +0.025$; $\tau$ $53 \to 54$ min &
  band $+0.35 \to +0.33$; anti-causal constant $-0.36 \to -0.39$; $\Aeven$ $+0.051 \to +0.055$ \\
\addlinespace
falls against rises & the two wings' decays fitted separately, per epoch where both resolve &
  $\tau_\downarrow/\tau_\uparrow = 1.02$ (0.84 to 1.23), 8 of 33 epochs &
  0.97 (0.81 to 1.16), 10 of 33 &
  no epoch resolves \\
\addlinespace
four-week epochs & the epochs cut to four weeks, 84 in place of 33 &
  $\Delta_s$ $0.071 \pm 0.008 \to 0.060 \pm 0.011$; pooled $\Delta$ $+0.182 \to +0.154 \pm 0.026$ &
  $\Delta_s$ $0.066 \pm 0.008 \to 0.076 \pm 0.010$; pooled $\Delta$ $+0.130 \to +0.134 \pm 0.018$ &
  not comparable: 29 of 84 epochs excluded at 20 min, 12 of 20 lags kept; $\Delta_s$ $0.026 \to 0.095 \pm 0.029$; anti-causal $p$ $0.005 \to 0.46$ \\
\addlinespace
window length & $W = 10$, 40 and 60 min beside 20, the lag grid in steps of $W$; the daily-cycle variants and the two Bitcoin nulls run at every window; the values in the order $W = 10$, 20, 40, 60 &
  half-decay lag 39, 89, 86, 120 min; $\Delta_s$ 0.044, 0.071, 0.074, 0.087; anti-causal constant over the first-lag value $-0.15$, $-0.03$, $-0.28$, $-0.38$; anti-causal $p$ 0.005, 0.46, 0.06, 0.98; $\Sodd/\Aeven$ at the first lag 0.5, 0.9, 0.4, 0.6 &
  35, 90, 100, 96 min; 0.054, 0.066, 0.086, 0.065; $-0.29$, $-0.09$, $-0.54$, $-1.4$; ${<}\,0.0025$, 0.58, 0.03, 0.03; 0.4, 0.7, 0.3, 0.0 &
  $W = 10$: anti-causal constant $-0.19 \pm 0.05$, first lag $+0.36 \pm 0.07$, $\Sodd/\Aeven$ at the first lag $+0.09$: the mirror stays; $W = 40$: $-0.33 \pm 0.34$, $+1.8 \pm 0.3$, $\Sodd/\Aeven$ $+0.77$, 32 epochs; $W = 60$: six epochs, not read; $W = 20$ with the instants every 40 min (the 40-minute sample): $\Delta_s$ $0.041 \pm 0.031$, anti-causal constant $-0.38 \pm 0.26$, anti-causal $p$ 0.92; the stocks at $W = 40$: $\Delta_s$ $0.057 \pm 0.026$ (Apple), $0.068 \pm 0.023$ (Tesla), $-0.008 \pm 0.027$ (Microsoft) against 0.027, 0.035 and 0.039 at 20 \\
\addlinespace
return spacing & the realized variance from one-minute returns in place of the two-second marks, the trend's marks unchanged; for Bitcoin and the index fund also from returns at 10 s, 30 s, 2 and 5 min &
  $\Delta_s$ $0.071 \to 0.087 \pm 0.011$ (per epoch $+0.012 \pm 0.005$); $\Aeven$ $+0.038 \to +0.048$; $\Sodd$ $+0.032 \to +0.034$; level $-0.04 \pm 0.03$; anti-causal $p$ $0.46 \to 0.22$; at 10 s, 30 s, 2 and 5 min: $\Delta_s$ 0.077, 0.084, 0.080, 0.076; $\Sodd$ $+0.032$, $+0.034$, $+0.033$, $+0.029$; the level follows the sampling bias of the logarithm at every spacing &
  $0.066 \to 0.076 \pm 0.009$ ($+0.014 \pm 0.004$); $\Aeven$ $+0.037 \to +0.044$; $\Sodd$ $+0.026 \to +0.033$; level $-0.04 \pm 0.02$; anti-causal $p$ $0.58 \to 0.38$ &
  $0.026 \to 0.042 \pm 0.017$ ($+0.025 \pm 0.010$); $\Aeven$ $+0.051 \to +0.073$; $\Sodd$ $-0.025 \to -0.022$; level $-0.19 \pm 0.02$; anti-causal $p$ $0.005 \to 0.005$: the mirror stays; at 10 s, 30 s, 2 and 5 min: $\Delta_s$ 0.037, 0.046, 0.047, 0.035; $\Sodd$ $-0.022$, $-0.020$, $-0.015$, $-0.026$; the level below the sampling bias by 0.08 at 10 s and by 0.12 to 0.16 from 30 s to 2 min \\
\addlinespace
wing fit & a quadratic term per wing (every asset); the trend cut at 4 and at 1.5 scales in place of 2.5 (the three primary assets); $\Delta_s$ at the first lag with the paired shift per epoch in brackets, the curvature $c_2 s^2$ per wing as its odd component $(c_2^- - c_2^+)/2$ and even component, the constant over the anti-causal lags in the unit of $\Delta$ &
  quadratic: $\Delta_s$ $0.071 \to 0.078 \pm 0.024$ ($+0.008 \pm 0.025$), curvature odd $-0.002 \pm 0.006$, even $+0.010 \pm 0.015$, constant $-0.005 \to +0.04 \pm 0.06$; cut 4: $0.074 \pm 0.009$ ($+0.002 \pm 0.005$), constant $+0.02 \pm 0.03$; cut 1.5: $0.071 \pm 0.012$ ($+0.004 \pm 0.011$), constant $0.00 \pm 0.03$ &
  quadratic: $0.066 \to 0.074 \pm 0.026$ ($+0.018 \pm 0.026$), odd $-0.004 \pm 0.006$, even $+0.048 \pm 0.011$, constant $-0.012 \to -0.04 \pm 0.06$; cut 4: $0.065 \pm 0.008$ ($-0.001 \pm 0.005$), constant $-0.006 \pm 0.014$; cut 1.5: $0.070 \pm 0.012$ ($+0.004 \pm 0.011$), constant $-0.025 \pm 0.034$ &
  quadratic: $0.026 \to 0.30 \pm 0.07$ ($+0.31 \pm 0.07$), odd $-0.071 \pm 0.017$, even $+0.021 \pm 0.026$, constant $-0.36 \to +0.73 \pm 0.40$, anti-causal $p$ $0.005 \to 0.24$, $\Sodd$ $-0.025 \to +0.21 \pm 0.06$; cut 4: $0.019 \pm 0.013$, constant $-0.26 \pm 0.09$, $p$ 0.005; cut 1.5: $0.110 \pm 0.026$, constant $-0.16 \pm 0.14$, $p$ 0.14, $\Sodd$ $+0.046 \pm 0.023$; the stocks with the quadratic term: Apple $0.027 \to 0.17 \pm 0.05$, Tesla $0.035 \to 0.13 \pm 0.05$, Microsoft $0.039 \to 0.09 \pm 0.06$, odd curvature $-0.03$, $-0.02$, $-0.02$ \\
\end{longtable}
\endgroup

\section{The three stocks}
\label{app:stocks}

\begin{figure}[tp]
\centering
\includegraphics[width=\textwidth]{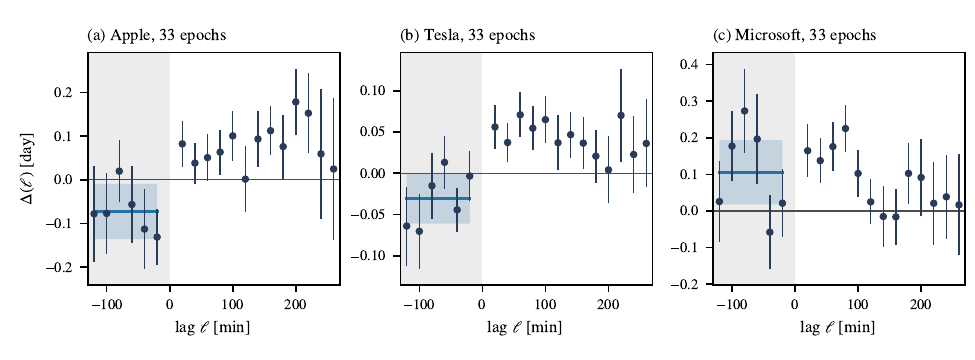}
\caption{The response of the three stocks versus lag. The asymmetry
$\Delta(\ell)$ in the log channel, the mean over 33 epochs with standard
error per lag, as in Fig.~\ref{fig:response}, for (a) Apple, (b) Tesla
and (c) Microsoft. The gray region denotes anti-causal lags; over them a
constant (blue) is plotted, computed as each epoch’s mean over its
anti-causal lags averaged across epochs with its standard error, as in
Fig.~\ref{fig:response}. Each curve terminates at the last lag retained
by the pool, 260 minutes, within the session. No fit is plotted over
causal lags: the time scale is reported for cryptocurrencies only
(Section~\ref{sec:time-scale}). Each panel uses its own vertical scale.}
\label{fig:response-stocks}
\end{figure}

\begin{figure}[tp]
\centering
\includegraphics[width=\textwidth]{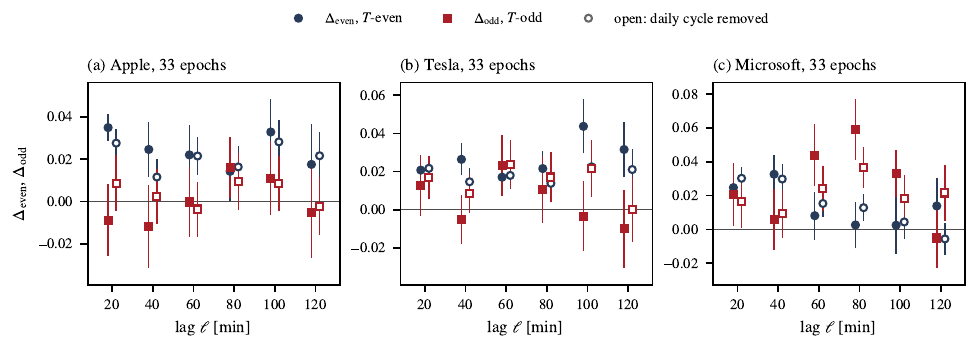}
\caption{Time reversal for the three stocks. The $T$-even part $\Aeven$
(circles) and the $T$-odd part $\Sodd$ (squares) of the scale-free
asymmetry $\Delta_s$, Eq.~(\ref{eq:time-reversal}) evaluated per epoch,
at the six lags from 20 to 120 minutes, averaged over 33 epochs with
standard error of the mean, for (a) Apple, (b) Tesla and (c) Microsoft.
Filled markers correspond to the values in
Table~\ref{tbl:time-reversal-stocks}; open markers, shifted slightly to
the right to prevent overlap of error bars, display the same quantities
after removal of the daily volatility cycle
(Section~\ref{sec:robustness}). Each panel uses its own vertical scale.}
\label{fig:time-reversal-stocks}
\end{figure}

Figures~\ref{fig:response-stocks} and~\ref{fig:time-reversal-stocks}
display, for Apple, Tesla and Microsoft, the response curve from
Fig.~\ref{fig:response} and the time-reversal parts from
Fig.~\ref{fig:time-reversal}, generated by the same scripts from the
same runs. Table~\ref{tbl:time-reversal-stocks} lists the parts as
Table~\ref{tbl:time-reversal} does for the other four assets. The
interpretation is provided in Section~\ref{sec:equities}: at 20 minutes
none of the three shows a detectable $T$-odd part and Microsoft’s values
are too noisy to resolve one across the band. Their constants over
anti-causal lags, $-0.07 \pm 0.06$, $-0.03 \pm 0.03$ and
$+0.11 \pm 0.09$, lie within roughly one standard error of zero and
differ in sign, consistent with their anti-causal tests
(Table~\ref{tbl:main}).

\begin{table}[htp]
\centering
\caption{Time reversal for the three stocks. The $T$-even part $\Aeven$
and the $T$-odd part $\Sodd$ of the scale-free asymmetry at the six lags
common to both sides, constructed per epoch and averaged over 33 epochs
with standard error of the mean, in units of $10^{-2}$; the final column
reports the ratio of the two band means over 20 to 120 minutes, the
median across epoch-bootstrap replicas with its 68\% interval, as in
Table~\ref{tbl:time-reversal}. A response absent before the trend
satisfies $\Sodd = \Aeven$; a mirror satisfies $\Sodd = 0$.}
\label{tbl:time-reversal-stocks}
\scriptsize
\setlength{\tabcolsep}{4pt}
\begin{tabular}{@{}llrrrrrrl@{}}
\toprule
 & & \multicolumn{6}{c}{lag $\ell$ [min]} & \\
\cmidrule(lr){3-8}
asset & & 20 & 40 & 60 & 80 & 100 & 120 & $\Sodd/\Aeven$, 20--120 min \\
\midrule
AAPL & $\Aeven$ & $+3.5 \pm 0.6$ & $+2.5 \pm 1.3$ & $+2.2 \pm 1.4$ & $+1.4 \pm 1.4$ & $+3.3 \pm 1.5$ & $+1.8 \pm 1.9$ & $+0.01$ [$-0.51$, +0.71] \\
     & $\Sodd$ & $-0.9 \pm 1.7$ & $-1.2 \pm 2.0$ & $-0.0 \pm 1.6$ & $+1.6 \pm 1.4$ & $+1.1 \pm 1.8$ & $-0.5 \pm 2.2$ & \\
TSLA & $\Aeven$ & $+2.1 \pm 0.6$ & $+2.6 \pm 0.8$ & $+1.7 \pm 1.0$ & $+2.2 \pm 0.9$ & $+4.4 \pm 1.4$ & $+3.2 \pm 1.5$ & $+0.17$ [$-0.35$, +0.78] \\
     & $\Sodd$ & $+1.3 \pm 1.6$ & $-0.5 \pm 1.3$ & $+2.3 \pm 1.6$ & $+1.1 \pm 1.7$ & $-0.3 \pm 1.8$ & $-1.0 \pm 2.1$ & \\
MSFT & $\Aeven$ & $+2.5 \pm 1.0$ & $+3.3 \pm 1.1$ & $+0.8 \pm 1.4$ & $+0.3 \pm 1.4$ & $+0.2 \pm 1.7$ & $+1.4 \pm 1.7$ & $+1.48$ [+0.28, +4.49] \\
     & $\Sodd$ & $+2.1 \pm 1.9$ & $+0.6 \pm 1.8$ & $+4.4 \pm 1.8$ & $+5.9 \pm 1.8$ & $+3.3 \pm 1.4$ & $-0.5 \pm 1.8$ & \\
\bottomrule
\end{tabular}
\end{table}

\clearpage

\section*{CRediT authorship contribution statement}
\textbf{Gülşah Bıçkı:} Data curation, Formal analysis, Investigation,
Visualization, Writing -- original draft. \textbf{Taylan Akdoğan:}
Conceptualization, Methodology, Software, Supervision, Writing -- review
\& editing.

\section*{Declaration of competing interest}
The authors declare that they have no known competing financial
interests or personal relationships that could have appeared to
influence the work reported in this paper.

\section*{Data availability}
The raw trade data are available from Massive.com (formerly Polygon.io)
under license. The derived data can be made available on request.

\section*{Acknowledgments}
The authors thank Massive.com, Inc.\ for providing the market data used
in this study free of charge. G.B.\ is supported by a graduate
scholarship from Özyeğin University. This research did not receive any
specific grant from funding agencies in the public, commercial, or
not-for-profit sectors.

\section*{Declaration of generative AI and AI-assisted technologies in the manuscript preparation process}
During the preparation of this work, the authors used Claude (Anthropic)
to proofread the manuscript for the fluency of its English. After using
this tool, the authors reviewed and edited the content as needed and
take full responsibility for the content of the published article.

\bibliographystyle{unsrtnat}

\bibliography{variance_response_arxiv}

\end{document}